\documentclass[useAMS,usenatbib]{mn2e}
\usepackage{graphicx,bm,amssymb,amsmath}
\usepackage{color}

\newcommand{\lcdm}{$\Lambda$CDM}

\newcommand{\OmMo}{\Omega_{\mathrm{M},0}}
\newcommand{\Ombo}{\Omega_{\mathrm{b},0}}

\newcommand{\OmLo}{\Omega_{\mathrm{\Lambda},0}}
\newcommand{\msol}{\mathrm{M}_{\odot}}
\newcommand{\sig}{\sigma_{8}}
\newcommand{\rvir}{\mathrm{R}_{\mathrm{vir}}}
\newcommand{\mvir}{M_{\rm vir}}
\newcommand{\smax}{s_{\mathrm{max}}}
\newcommand{\EinRad}{\theta_{\mathrm{E}}}

\newcommand{\mincir}{\raise
  -2.truept\hbox{\rlap{\hbox{$\sim$}}\raise5.truept \hbox{$<$}\ }}
\newcommand{\magcir}{\raise
  -2.truept\hbox{\rlap{\hbox{$\sim$}}\raise5.truept \hbox{$>$}\ }}
\newcommand{\siml}{\raise
  -2.truept\hbox{\rlap{\hbox{$\sim$}}\raise5.truept \hbox{$<$}\ }}
\newcommand{\simg}{\raise  -2.truept\hbox{\rlap{\hbox{$\sim$}}\raise5.truept \hbox{$>$}\ }}

\title[Baryons \& Cluster Strong Lensing]{How Baryonic Processes affect Strong Lensing properties of Simulated Galaxy Clusters}

\author[Killedar et al.]{
M.~Killedar$^{1,2}$\thanks{
E-mail: killedar@oats.inaf.it (MK)
}, 
S. Borgani$^{1,2,3}$, 
M. Meneghetti$^{4,5}$,
K. Dolag$^{6,7}$, 
D. Fabjan$^{1,8,9}$\\ ~\\
\LARGE{\rm 
\& L. Tornatore$^{1}$
} 
\\~\\
$^1$ Dipartimento di Fisica dell'Universit\`{a} di Trieste, Sezione di Astronomia, Via Tiepolo 11, I-34131 Trieste, Italy\\
$^2$ INAF - Osservatorio Astronomico di Trieste, Via G.B. Tiepolo 11, I-34131 Trieste, Italy\\
$^3$ INFN - National Institute for Nuclear Physics, Via Valerio 2, I-34127 Trieste, Italy\\
$^4$ INAF - Osservatorio Astronomico di Bologna, Via Ranzani 1, I-40127 Bologna, Italy\\
$^5$ INFN - Sezione di Bologna, Viale Berti Pichat 6/2, I-40127 Bologna, Italy\\
$^6$ Universit\"atssternwarte M\"unchen, M\"unchen, Germany\\
$^7$ Max-Planck-Institut f\"ur Astrophysik, Garching, Germany\\
$^8$ SPACE-SI, Slovenian Centre of Excellence for Space Sciences and 
Technologies, A$\check{s}$ker$\check{c}$eva 12, 1000 Ljubljana, Slovenia \\
$^9$ Faculty of Mathematics and Physics, University of Ljubljana,
Jadranska 19, 1000 Ljubljana, Slovenia\\ 
}

\begin{document}
\bibliographystyle{scemnras} 

\date{Accepted 2012 August 23. Received 2012 August 23; in original form 2012 August 03}

\pagerange{\pageref{firstpage}--\pageref{lastpage}} \pubyear{2011}

\maketitle

\label{firstpage}


\begin{abstract}
  The observed abundance of giant arcs produced by galaxy cluster
  lenses and the measured Einstein radii have presented a source of
  tension for \lcdm, particularly at low redshifts
  ($z\sim0.2$). Previous cosmological tests for high-redshift clusters
  ($z>0.5$) have suffered from small number statistics in the
  simulated sample and the implementation of baryonic physics is
  likely to affect the outcome. We analyse zoomed-in simulations of a
  fairly large sample of cluster-sized objects, with
  $\mvir>3\times10^{14}\,h^{-1}\msol$, identified at $z=0.25$ and $z=0.5$, for
  a concordance $\Lambda$CDM cosmology. These simulations have been carried out by incrementally increasing the physics considered. We start with dark matter only simulations, and then add gas hydrodynamics, with different treatments of baryonic processes: nonÐradiative cooling, 
radiative cooling with star formation and galactic winds powered by supernova explosions, and finally including the effect of AGN feedback. Our analysis of strong lensing properties is based on
  the compuation of the cross-section for the formation of giant arcs
  and of the Einstein radii. We find that the addition of gas in
  non--radiative simulations does not change the strong lensing
  predictions significantly, but gas cooling and star formation
  together significantly increase the number of expected giant arcs
  and the Einstein radii, particularly for lower redshift clusters
  {\it and} lower source redshifts. Further inclusion of AGN feedback
  reduces the predicted strong lensing efficiencies such that the
  lensing probability distributions becomes closer to those
  obtained for simulations including only dark matter. Our results
  indicate that the inclusion of baryonic physics in simulations will
  not solve the arc-statistics problem at low redshifts, when the
  physical processes included provide a realistic description of
  cooling in the central regions of galaxy clusters. As outcomes of
  our analysis, we encourage the adoption of Einstein radii as a
  robust measure of strong lensing efficiency, and provide the
  \lcdm~predictions to be used for future comparisons with
  high-redshift cluster samples.
\end{abstract}


\section{Introduction}
The hierarchical formation paradigm describes a scenario in which
small structures form earlier and then merge to form more massive
objects. This description within the \lcdm~model has been highly
successful at explaining observations of structure on a large range of
scales and redshifts \citep[e.g.][]{ESM90,WF91,K11}, including the
formation of galaxy clusters \citep[e.g.][]{KB12}. However, the
internal structure of galaxy clusters has posed a challenge. The
earliest comparisons between simulated clusters and the observed
frequency of gravitational lensing arcs revealed a serious discrepancy
between the observations and \lcdm~predictions \citep{B98,L05}. More
recent comparisons to X-ray selected clusters find that the
discrepancy remains for $z<0.3$ and $z>0.5$ but results are not
definitive due to small simulated sample sizes at higher redshifts and
uncertainties with selection procedures and biases \citep{M11,H11}. In
addition, observations of clusters at redshifts $0.1<z<0.6$ have found
higher concentrations and larger Einstein radii than predicted
\citep{BB08, SR08, Z11a}. These are related problems since both
suggest a failure of \lcdm~to correctly infer the distribution of
matter in the cores of clusters; the solution may be found in the
details of the simulations used to formulate the predictions.

Simulations which only incorporate collisionless particles are
commonplace in the literature. Dark matter structure is evolved under
gravity with the effect of baryons ignored. On large-scales, the
distribution of gas follows the dark matter potential wells, so the
collisionless simulations will suffice; however on galaxy- and group-
scales, gas cooling, star formation and feedback can alter the
gravitational potential significantly
\citep[e.g.][]{BFFP86,G04}. Early simulations that were used to model
the cluster lenses did not incorporate sufficient relevant physical
processes \citep{HWY07,SR08}. These simulations were originally
justified by the premise that the temperature of the ICM is too hot to
allow efficient cooling, so the baryonic matter must simply follow the
dark matter potential and would not have a significant impact on its
shape \citep[e.g.][]{BB08}. The role of baryons in shaping galaxy
clusters has recently become a key point of discussion since
hydrodynamic simulations have become more feasible; the findings are
that baryons do, in fact, affect the shapes of gravitational potential
of the simulated haloes. Gas cooling, for example, leads to adiabatic
compression of galaxy haloes preferentially in the cluster centre,
allowing more mass to be deposited around the core
\citep{BL10}. Cooling and star formation serve to steepen the density
profile of clusters and thus increase the mass concentration and the
strong lensing cross-section \citep{Lewis00,G04,P05,R08,RZK08,Cui12}.
Unfortunately, these processes lead to the well known overcooling
problem, in which the cold gas mass and stellar mass in the cluster
core is overestimated.

Including in simulations feedback from gas accretion onto supermassive
black holes (SMBHs) is a promising solution for overcooling
\citep{Sijacki07,T11,McC10,Fabjan10}, while simultaneously
reproducing the drop in the cosmic SFR for $z<2$
\citep[e.g.,][]{vdV11}. The resulting feedback from active galactic
nuclei (AGN) significantly reduces the gas fraction in simulated
galaxy-groups and poor clusters ($T\lesssim2$keV), but cannot remove
gas from the deep potential wells of rich, massive clusters
\citep{PSS08,Fabjan10,McC10}. While the total baryonic content of
massive clusters is relatively unaffected by cooling, star formation
and energy feedback, such processes can imply a quite pronounced
redistribution of baryons, particularly close to the cluster
centre. \citet{D10} analysed simulated haloes at $z=0$ ranging from
galaxy to cluster scales. They demonstrated that while radiative
cooling increases mass concentration\citep[see also][]{RZK08},
AGN feedback has the opposite effect; clusters modelled with AGN
feedback have mass concentrations equal to those modelled with dark
matter only. Consistent with these results, \citet{Mead10}
demonstrated that AGN feedback is able to reduce the strong lensing
cross section for clusters at $z=0.2$. Four of the five clusters they
simulated had strong lensing efficiencies consistent with their
dark-matter counterparts, while the fifth remained a stronger lens.

Many of the previous works investigating the role of baryons in
modifying cluster cores have told similar stories: while cooling and
star-formation either directly or indirectly leads to an increase of
mass in the core, and therefore an increase in the strong lensing
efficiency, AGN feedback negates this to some extent. Unfortunately
previous studies have had limited samples of massive clusters,
especially at high-redshift $z\gtrsim0.3$, and have utilised
simulations generated with values for cosmological parameters that
have since been revised. The strong lensing studies of \citet{P05}, \citet{R08} and
\citet{Mead10} were undertaken in the WMAP-1 type cosmology with
$\sig=0.9$. As shown by \citet{M08}, changing the
cosmology from WMAP-1 to WMAP-3 best fitting models can lead to a 20\%
decrease in the predicted mass concentration of relaxed clusters. This
is primarily due to the decrease of the power-spectrum normalisation,
$\sig$, from 0.9 to 0.7, and the subsequent delay in the assembly of
haloes. The current WMAP-7 best-fit values lie between these two
extremes; the expected reduction in the concentration of masses considered in this work from WMAP1 to WMAP5/7 is closer to 14\% \citep[e.g.][but see also \citealt{Prada12}]{ M08, D08}.

In this work we improve upon the aforementioned works by analysing the
strong lensing efficiencies of a larger sample of clusters simulated
within the currently favoured cosmology ($\sig=0.8$) with only dark
matter (DM), and also including the hydrodynamical treatment of gas,
along with cooling, star formation and feedback from both supernova
(SN) and AGN. Strong lensing properties of our fairly large sample of
massive clusters are analysed at two redshifts, $z=0.25$ and 0.5. A
description of the cosmological simulations and the cluster sample
follows in Section \ref{sims}. Earlier studies have characterised
strong lensing by the cross section for the formation of giant arcs. A
comparison with observational data requires the cross-section, as a
function of source redshift, to be convolved with the source redshift
distribution; uncertainties in this source redshift distribution
weaken the cosmological test. We, therefore, propose that the Einstein
radius is a more robust proxy for strong lensing; critical curves are
determined at a single source redshift (usually $z_{s}=2$), regardless
of the range of source redshifts used for lens mass
reconstructions. Therefore, in Section \ref{Xsec} and Section \ref{ER}
we present the strong-lensing properties for the relaxed sub-sample
and discuss the influence of the baryonic processes; in the former
section we characterise strong-lensing with the cross-section for the
formation of giant-arcs, while in the latter we consider the Einstein
Radii instead. At the end of this section, we describe the
strong-lensing properties of unrelaxed sub-sample. These results are
then interpreted in terms of variation of the density profiles induced
by the presence of baryons in the simulated clusters. Finally, we
summarise of our findings in Section~\ref{conclude}.

\section{Simulations}\label{sims}
\subsection{The simulated set of clusters}
The set of simulated clusters analysed in this study have been
previously presented by \citep{Fabjan10} and \citet{B11}, and a more
comprehensive description will be provided in a forthcoming paper
(Planelles et al. in prep), but we describe them in brief here.
Cluster halos have been identified in a low--resolution simulation box
having a periodic co-moving size of 1~h$^{-1}$ Gpc for a flat \lcdm~
model whose cosmological parameters were chosen as follows: present
day vacuum density parameter, $\OmLo=0.76$; matter density parameter,
$\OmMo=0.24$; baryon density parameter, $\Ombo=0.04$; Hubble constant
$h=0.72$; normalisation of the matter power spectrum $\sig=0.8$; and
primordial power spectrum $P(k) \propto k^{n}$ with $n=0.96$. The
parent simulation followed 1024$^{3}$ collision-less particles in the
box. Clusters were identified at $z=0$ using a standard {\it
  Friends-of-Friends} (FoF) algorithm, and Lagrangian regions around
24 of the clusters found to have masses
$M_{\mathrm{FOF}}>10^{15}$h$^{-1}\msol$ were re-simulated at higher
resolution employing the {\it Zoomed Initial Conditions} code
\citep[ZIC;][]{TBW97}, while resolution is progressively degraded
outside these regions, so as to save computational time while still
providing a correct description of the large--scale tidal field. The
Lagrangian regions were large enough to ensure that only
high-resolution particles are present within five virial-radii of the
central cluster.

Simulations have been carried out using the TreePM--SPH {\small
  GADGET--3} code, a newer version of the original {\small GADGET--2}
code by \cite{S05} that adopted a more efficient domain
decomposition to improve the work-load balance. Each re-simulation has
been repeated with a unique set of included baryonic processes, some
of which are analysed here: one simulation-set uses collisionless dark
matter particles only; one is a non-radiative run; another two
implement a number of processes associated with baryons; one of these
two includes AGN feedback. The basic characteristics of these
re-simulation sets are described here below.

\begin{description}
\item[{\bf DM}]: simulations including only dark matter particles,
  that in the high-resolution region have a mass
  $m_{DM}=10^{9}h^{-1}\msol$. The Plummer--equivalent co-moving
  softening length for gravitational force in the high--resolution
  region is fixed to $\epsilon_{Pl}=5 h^{-1}$ kpc physical at $z<2$
  while being fixed to $\epsilon_{Pl} = 15
  h^{-1}$ kpc comoving at higher redshift.\\
\item[{\bf NR}]: non--radiative hydrodynamical simulations. Initial
  conditions for these hydrodynamical simulations are generated
  starting from those of the DM-only simulations, and splitting each
  particles in the high resolution region into one dark matter and one
  gas particle, with their masses chosen so to reproduce the assumed
  cosmic baryon fraction.  The mass of each DM particle is then
  m$_{\rm{DM}} = 8.47 \cdot 10^8 \, \rm{h}^{-1} \msol$ and the mass of
  each gas particle is m$_{\rm{gas}} = 1.53 \cdot 10^8 \, \rm{h}^{-1}
  \msol$. For the computation of the hydrodynamical forces we assume
  the minimum value attainable by the SPH smoothing length of the
  B--spline interpolating kernel to be half of the corresponding value
  of the gravitational softening length.  No radiative cooling is
  involved.
\item[{\bf CSF}]: hydrodynamical simulations including the effect of
  cooling, star formation and SN feedback. Radiative cooling rates are
  computed by following the same procedure presented by
  \cite{WSS09}. We account for the the presence of the cosmic
  microwave background (CMB) and for the model of UV/X--ray background
  radiation from quasars and galaxies, as computed by
  \cite{haardt_madau01}. Contributions to cooling from each one of
  eleven elements (H, He, C, C, N, O, Ne, Mg, Si, S, Ca, Fe) have been
  pre--computed using the publicly available {\footnotesize {\sc
      CLOUDY}} photo--ionisation code \citep{ferland_etal98} for an
  optically thin gas in (photo--)ionisation equilibrium. Gas particles
  above a given threshold density are treated as multiphase, so as to
  provide a sub-resolution description of the inter-stellar medium,
  according to the model originally described by
  \cite{SH03}. Within each multiphase gas particle, a
  cold and a hot-phase coexist in pressure equilibrium, with the cold
  phase providing the reservoir of star formation.  We include a
  detailed description of metal production contributed by SN-II, SN-Ia
  and low and intermediate mass stars, as described by
  \cite{T07}. Stars of different mass, distributed
  according to a Chabrier IMF \citep{chabrier03}, release metals over
  the time-scale determined by the corresponding mass-dependent
  life-times (taken from \citealt{PM93}).  Kinetic
  feedback contributed by SN-II is implemented according to the scheme
  introduced by \cite{SH03}: a multi-phase star
  particle is assigned a probability to be uploaded in galactic
  outflows, which is proportional to its star formation rate. For this
  set of simulations we assume $\rm{v}_{\rm{w}} = 500$ km s$^{-1}$ for the
  wind velocity.
\item[{\bf AGN}]: the same as CSF, with a lower wind velocity of
  $\rm{v}_{\rm{w}} = 350$ km s$^{-1}$, also including the effect of AGN
  feedback.  In the model for AGN feedback, released energy results
  from gas accretion onto SMBHs.  This model introduces some
  modifications with respect to that originally presented by
  \cite{SdMH05} (SMH hereafter), to which is largely
  inspired, and will be described in detail by Dolag et al. (2012, in
  preparation). BHs are described as sink particles, which grow their
  mass by gas accretion and merging with other BHs. Gas accretion
  proceeds at a Bondi rate, while being Eddington--limited. Once the
  accretion rate is computed for each BH particle, a stochastic
  criterion is used to decide which of the surrounding gas particles
  contribute to the accretion. Unlike in SMH, in which a selected gas
  particle contributes to accretion with all its mass, we included the
  possibility for a gas particle to accrete only with a slice of its
  mass, which corresponds to 1/4 of its original mass. In this way,
  each gas particle can contribute with up to four ``generations'' of
  BH accretion events, thus providing a more continuous description of
  the accretion process \citep[see also][]{Fabjan10}.  BH
  particles are initially seeded with a mass of $0.05 \,
  \rm{m}_{\rm{DM,}10}$, where $\rm{m}_{\rm{DM,}10}$ is the DM particle
  mass in units of $10^{10} \rm{h}^{-1}M_\odot$. Seeding of BH
  particles takes place in halos when they first reach a minimum
  friend-of-friend (FoF) mass of $2.5\times 10^3 \,
  \rm{m}_{\rm{DM,}10}$ (using a linking length of 0.16 in units of the
  mean interparticle separation in the high-resolution region), with
  the further condition that such halos should contain a minimum mass
  fraction in stars of 0.02. The first condition on the minimum halo
  mass guarantees that such halos are resolved with at least $\magcir
  200$ DM particles, while the second condition requires that
  substantial star formation took place in such halos. This criterion
  prevents seeding BHs in halos possibly located at the border of the
  high resolution region, which spuriously contain a low amount of
  cooled gas, due to the interaction with nearby low--resolution DM
  particles.  Eddington-limited Bondi accretion produces a radiated
  energy which correspond to a fraction $\epsilon_r= 0.1$ of the
  rest-mass energy of the accreted gas, which is determined by the
  radiation efficiency parameter $\epsilon_r$. The BH mass is
  correspondingly decreased by this amount. A fraction of this
  radiated energy is thermally coupled to the surrounding gas. We use
  $\epsilon_f = 0.05$ for this feedback efficiency, which increases to
  $\epsilon_f = 0.2$ whenever accretion enters in the quiescent
  ``radio'' mode and takes place at a rate smaller than one-hundredth
  of the Eddington limit \citep[e.g.][]{Sijacki07,Fabjan10}.
\end{description}

\subsection{The identification of clusters}
The cluster haloes are identified as follows. Firstly, a standard {\it
  Friends-of-Friends} algorithm is run over the DM particles in the
high--resolution regions, using a linking length of 0.16 in units of
the mean-interparticle separation. Within each FoF group, we identify
the position of the particle with the minimum gravitational potential,
which is then taken as the centre from where clusters are then
identified according to a spherical overdensity (SO) method. The
virial radius is defined as the smallest radius of a sphere centred on
the cluster, for which the mean density falls below the virial
overdensity. The virial overdensity is measured relative to the
critical density and calculated using the fitting formula of
\citet{BN98}, so for clusters at $z=0.25$, $\Delta_{c}\approx112$ and
for clusters at $z=0.5$, $\Delta_{c}\approx129$.

\begin{figure}
	 \begin{center}  
      	\includegraphics[width=0.98\linewidth]{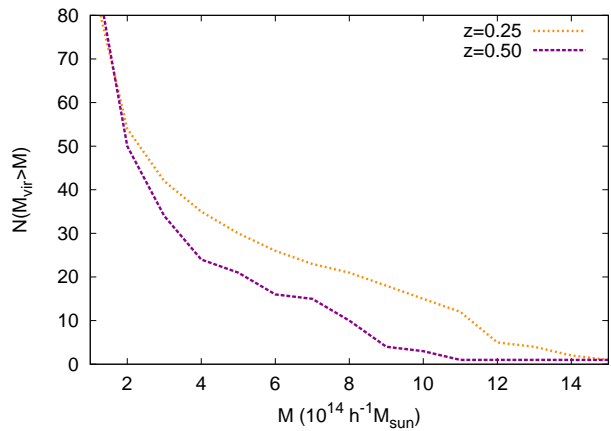}
      	\end{center}
      	\caption{The cumulative virial-mass distribution of clusters
          in the sample at redshifts $z=0.25$ (orange dotted curve)
          and $z=0.5$ (purple dashed curve).}
	\label{NumCounts}
\end{figure}
Cluster haloes are chosen for our strong lensing analysis only if no
low--resolution particles contaminate the region within five virial
radii of the cluster centre. In Figure~\ref{NumCounts} we show the
number of clusters above a given mass limit at the two redshifts
considered in our analysis. Note that this is not equivalent to a halo
mass function since the re-simulated regions do not cover the entire
box. The sample has 42 clusters with $\mvir>3\times10^{14}\,h^{-1}\msol$ at
a redshift of $z=0.25$. Moving to higher redshifts, we find 34
clusters with $\mvir>3\times10^{14}\,h^{-1}\msol$ at a redshift of $z=0.5$.

\subsection{Relaxed clusters}
If mergers between clusters take place along the line of sight
the strong lensing efficiency can be enhanced. If the
merger occurs across the sky, we will see multiple critical curves,
which positively biases the cross section for giant arcs. In the case
that critical curves are merging, both the Einstein radii and the
cross section for giant arcs would be biased large. This could confuse
our interpretation of the effects of baryon physics.

In order to characterise the degree of relaxation of each cluster, we
compute the quantity $\smax=\{max(s_{\zeta}): 0.05<\zeta<2\}$, where
\begin{equation}
	s_{\zeta} = \frac{\lvert \bm{r}_{\mathrm{COM}}(<\zeta \rvir)-\bm{r}_{\mathrm{MinPot}}\rvert}{\zeta \rvir}\,.
\end{equation}
In the above equation, $\bm{r}_{\mathrm{MinPot}}$ is the position of
the particle with the minimum gravitational potential and
$\bm{r}_{\mathrm{COM}}$ is the centre of mass of all the matter within
$\zeta \rvir$. In this way, $s_{\zeta}$ represents the offset between these
two measures of ``centre'' for any aperture radii, $\zeta
\rvir$, so that $\smax$ is the maximum offset over a range of aperture radii,
which are found by allowing $\zeta$ to vary from 0.05 to 2 in 30
logarithmic steps. We define a relaxed cluster as that for which the
counterpart in the {\tt DM} simulation has a maximum offset parameter
of $s<0.1$.
%
\begin{figure}
	 \begin{center}  
      	\includegraphics[angle=-90, width=0.99\linewidth]{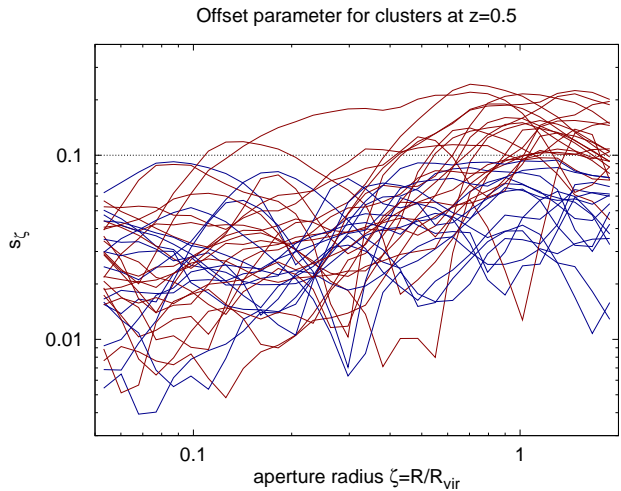}
      	\end{center}
      	\caption{The offset parameter, $s_{\zeta}$, as a function of
          aperture radius within which centre-of-mass is calculated,
          for all clusters at $z=0.5$. The dotted horizontal line
          shows the value of the maximum offset parameter,
          $\smax=0.1$, allowed for a cluster to be classified as
          relaxed. Red and blue curves correspond to clusters which
          are then classified as unrelaxed and relaxed, respectively.}
	\label{Sep1}
\end{figure}

As a visual aid, the offset parameter for each cluster at $z=0.5$ is
shown in Figure~\ref{Sep1} plotted against the aperture radius; each
curve corresponds to an individual cluster. Any cluster for which the
curve exceeds the dotted line ($s_{\zeta}=0.1$) for any radius in the
given range is considered to be unrelaxed.
Our adopted criterion to classify relaxed and unrelaxed clusters
corresponds to similar definitions in the literature when we choose
only one aperture radius, the virial radius i.e. when we set $\zeta=1$
\citep[e.g.][]{CER96, T98,N07,DON07,P11x}. However, we note that
adopting this common choice we would be less sensitive to complex mass
distributions near the core of a cluster, which is more important in
the context of strong lensing. Our definition, in a sense, weights the
disturbance caused by substructure by the inverse of radial position
of the substructure.
%
%
\begin{figure}
	 \begin{center}
      	\includegraphics[trim=2mm 3mm 2mm 0mm, clip, width=0.99\linewidth]{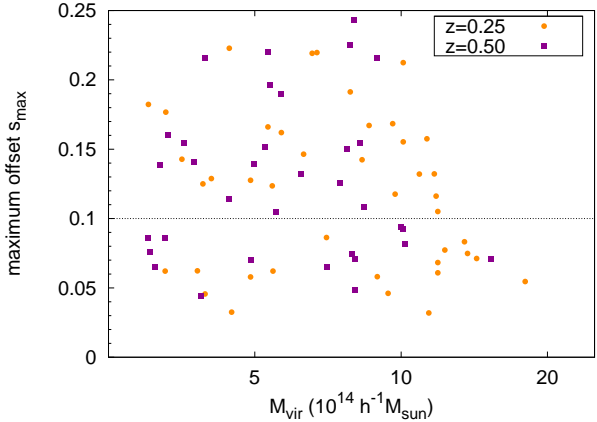}
      	\end{center}
      	\caption{The relation between the maximum centre shift,
          $\smax$, and the virial mass, $\mvir$, for all clusters. The
          orange circles represent clusters at $z=0.25$ while the
          purple squares represent the clusters at $z=0.5$.}
	\label{SmaxVsMvir}
\end{figure}
In order to verify whether there is any correlation within our sample
between degree of relaxation and cluster mass, we plot in
Figure~\ref{SmaxVsMvir} the positions of our clusters in the 
$\smax$-$\mvir$ plane. There is negligible correlation between the
cluster's mass and the offset parameter that describes the degree of 
relaxation. The choice of threshold for $\smax$ above which a cluster
is considered unrelaxed does not appear to bias the sample in terms of
mass.
The total sample of 42 clusters at $z=0.25$ includes 17 relaxed
clusters, and of the 34 clusters in total at $z=0.5$, 14 are
classified as relaxed. In Section~\ref{Xsec} and at the beginning of Section~\ref{ER} we
present the statistical results only for the relaxed sub-sample. In
Section~\ref{unrelaxed} we compare these results to those for unrelaxed
clusters. For comparison, we note that if we had defined the relaxed
sub-sample as those for which the offset at the virial radius is
$s_{1}<0.07$, as is common in the literature, we would have 23 relaxed
clusters at $z=0.25$ and 12 relaxed clusters at $z=0.5$.

\section{Cross Section for Giant Arcs}\label{Xsec}
The cross section for giant arcs is defined as the area in the source
plane in which if a distant galaxy (at that redshift) was located, it
would appear as a highly elongated giant arc due to the effects of
strong gravitational lensing. Every line of sight through each
simulated cluster will, in general, provide a unique cross section. We
now describe the procedure used to calculate the giant arc cross
section for each line of sight analysed for each cluster in our
simulated sample.
Throughout the present work, we refer to gravitational lensing
quantities following the notation of \citet{SEF92}. The lensing mass
is chosen to be contributed by all particles within two virial radii
of the cluster centre. We apply the thin lens approximation: the
assumption that the lensing mass is constrained to a single
plane. This treatment is valid for the present study, since the sizes
of the cluster lenses are much smaller than $D_{s}$, $D_{d}$ and
$D_{ds}$: the angular diameter distances from the observer to the
source, from the observer to the lens, and from the lens to the source,
respectively. The convergence, $\kappa$, can be defined by:
\begin{equation}\label{kappa}
\kappa = \frac{\Sigma}{\Sigma_{\mathrm{crit}}},
\end{equation}
where ${\Sigma}$ is the surface density and the critical surface
density for gravitational lensing is given by
\begin{equation}\label{sigmacrit}
\Sigma _{\mathrm{crit}} = \frac{c^{2}}{4\pi G}\frac{D_{s}}{D_{d}D_{ds}}.
\end{equation}
Note that from here onwards, the redshift of the cluster lenses is
denoted $z_{L}$, while the redshift of the background source galaxies
is denoted $z_{s}$.

\subsection{Calculating the lensing cross section}\label{findXsec}
In order to calculate the lensing cross-section, we first measure the
deflection of light-rays from background source galaxies across a
field of view in which we expect to search for giant arcs. Each ray of
light is deflected at the lens plane; the deflection angle is related
to the convergence by:
\begin{equation}
	\boldsymbol{\alpha} (\mathbf{x}) = \frac{1}{\pi}\int_{\mathbb{R}^{2}} d^2x'\kappa(\mathbf{x}')
			 \frac{\mathbf{x}-\mathbf{x'}}{\left|\mathbf{x}-\mathbf{x'}\right|^2},
	\label{defangle}
\end{equation}
where $\mathbf{x}$ is a dimensionless position in the lens plane. It is possible to choose the scaling such that the gravitational lens equation is given by:
\begin{equation}\label{lenseqn}
	\boldsymbol{\beta} 
	= \boldsymbol{\theta} - \boldsymbol{\alpha}(\boldsymbol{\theta}),
\end{equation}
where $\bm{\beta}$ is the angular source position and $\bm{\theta}$ is the angular position of the image on the sky. 
%
\begin{figure*}
      	\includegraphics[width=0.98\linewidth]{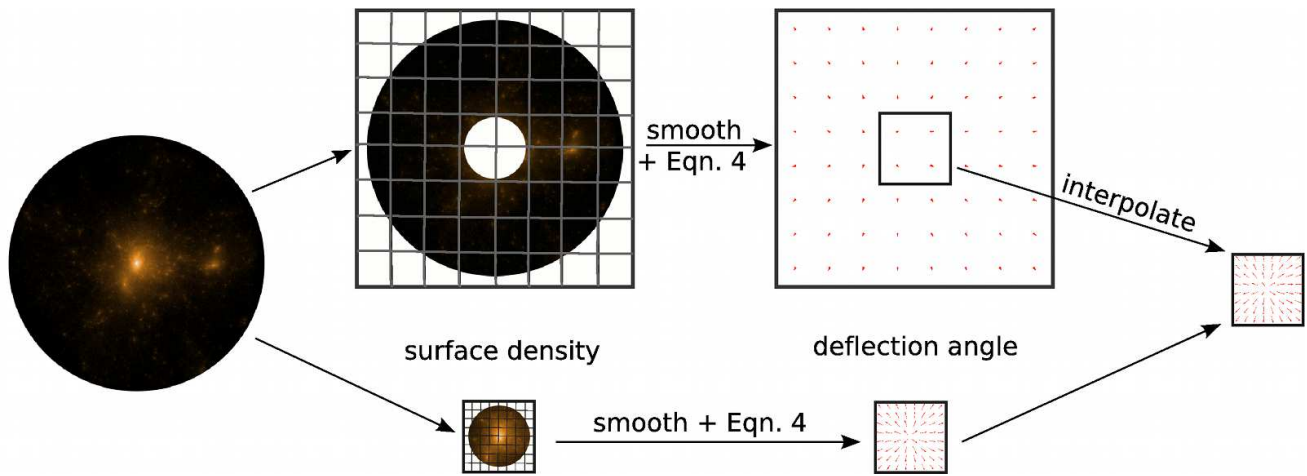}
      	\caption{A cartoon diagram describing the calculation of
          deflection angle, $\boldsymbol{\alpha}$, across a high
          resolution grid. The lensing mass is divided into two parts:
          an inner and an outer region (shown in the bottom and top rows,
          respectively), each of which are projected onto a small
          high-resolution and large low-resolution grid,
          respectively. The surface density distribution on each grid,
          $\kappa$, is smoothed, then the deflection angle is
          calculated using equation~\ref{defangle}. The total
          deflection angle is the sum of these two. However the
          deflection angle calculated in the inner region of the large
          grid must first be interpolated onto the small
          high-resolution grid. See text for further details}
	\label{DefMapHowTo}
\end{figure*}
%
Producing maps of the deflection angle at each grid point requires a few steps, as shown in the cartoon diagram in Figure~\ref{DefMapHowTo}. We describe the procedure below. The angular and spatial resolutions are quoted for sources at $z_{s}=2$ and clusters at $z_{L}=0.25$, but quantities for clusters at $z_{L}=0.5$ are given in square brackets. 
The lensing mass is projected onto one of two possible lens planes, depending on the position of the particles. 
Those that lie at a projected co-moving distance of more than 0.9 h$^{-1}$Mpc [1.35 h$^{-1}$Mpc] from the cluster centre are placed on a low-resolution $2048\times2048$ grid. The angular resolution of this grid is 1.7 arcsec [0.9 arcsec]. 
Particles that are projected closer to the cluster core, and are therefore responsible for the bulk of the strong lensing, are placed on a high-resolution $2048\times2048$ grid. The angular resolution of this grid is 0.3 arcsec [0.2 arcsec]. 
The mass on each grid is smoothed with a truncated Gaussian filter of size $\sigma=5$ h$^{-1}$kpc, to match the force-softening length of the simulation. The size of each grid is sufficiently larger than the projected extent of the mass on that grid, so that smoothing does not result in any loss of mass.

We convolve each of the two convergence maps with the appropriate
kernel from equation~\ref{defangle}.  This leaves us with two maps of
the deflection angle: one at high-resolution where the deflection is
solely due to the mass in the projected cluster core; the other at
low-resolution from the rest of the lensing mass.  For the subsequent
ray-shooting procedure, we use a single map of deflection angle for
each cluster projection; the final map has the same size and
resolution as the high-resolution grid described above. The deflection
angle at each grid point is determined with bilinear interpolation
from the closest grid-points of the low-resolution map, and adding the
corresponding grid-point from the high-resolution map. The resulting
deflection angle map has an angular resolution of
$0.\mspace{-5.0mu}"\mspace{-1.5mu}3$ for clusters at $z_{L}=0.25$ and
$0.\mspace{-5.0mu}"\mspace{-1.5mu}2$ for clusters at $z_{L}=0.5$.

Calculating the cross-section for giant arc formation requires a
ray-shooting and arc-identifying procedure, which has been previously
presented in \citet{M00} and \citet{M05}. Elliptical sources with an
equivalent radius of size $0.\mspace{-5.0mu}"\mspace{-1.5mu}5$ are
placed throughout the source plane; more sources are placed in regions
identified as caustics. The deflection angle maps allow mock images to
be generated from the sources. The images are fitted to ellipses and
the length-to-width ratio, $L/W$, is determined. If $L/W$ surpasses
some threshold elongation, $\eta$, the image is identified as a `giant
arc'. The cross-section for giant arc formation, $\sigma_{\eta}$, is
the area in the source plane covered by sources that are mapped onto
giant arcs. For the present study, the length-to-width threshold is
chosen to be $\eta=7.5$ --- following \citet{P05}, \citet{Mead10} and
\citet{M11} --- because much larger cross-sections are subject to
small number statistics, and much smaller cross-sections are too
sensitive to the intrinsic ellipticities of the source galaxy. We
discuss other choices of threshold of Section~\ref{baryonsXsec}.
 
We present results in this section and Section~\ref{ER} for sources at
redshift $z_{s}=2$, but include the statistical results for a source
redshift of $z_{s}=1$ in Table 1. For these lower
source redshifts, we would expect smaller Einstein radii, so the
deflection angle maps have a higher angular resolution:
$0.\mspace{-5.0mu}"\mspace{-1.5mu}2$ for clusters at $z_{L}=0.25$ and
$0.\mspace{-5.0mu}"\mspace{-1.5mu}1$ for clusters at $z_{L}=0.5$.

\subsection{The impact of baryons on the cross section for giant arcs}\label{baryonsXsec}
\begin{figure*}
      	\includegraphics[width=0.98\linewidth]{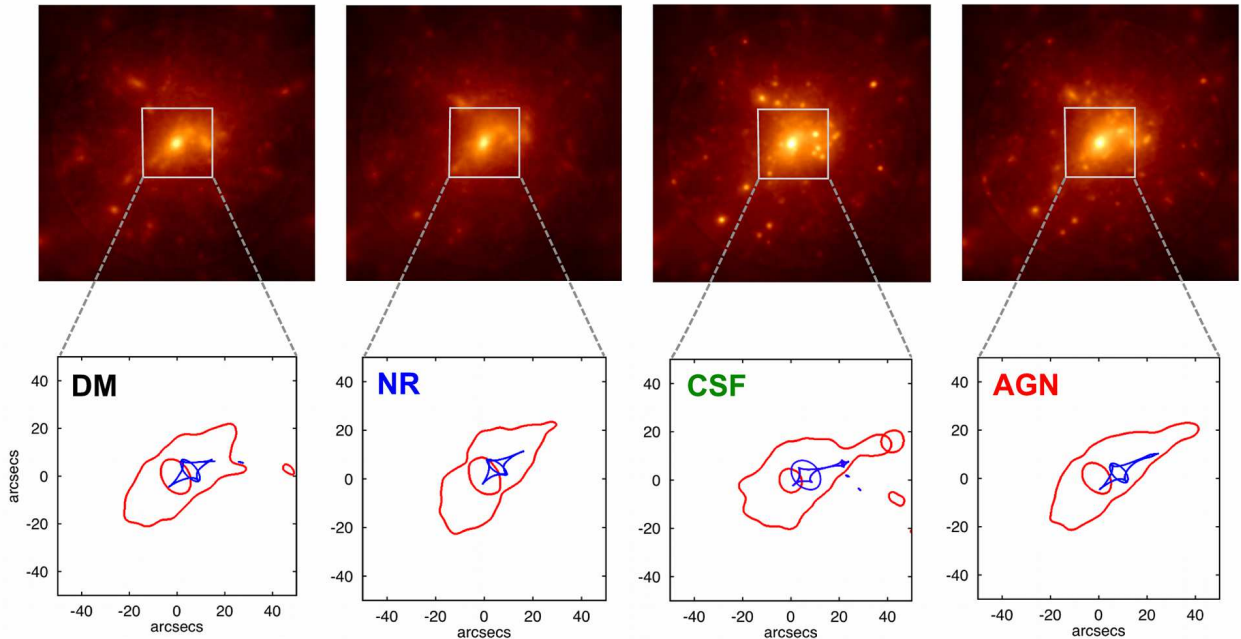}
      	\caption{One cluster at $z_{L}=0.5$ seen from the same viewpoint within
          different simulations: {\tt DM} (left panel), {\tt NR}
          (second panel), {\tt CSF} (third panel), and {\tt AGN}
          (right panel). Along the top row, we show the convergence
          map of the cluster, $\sim410$ arcsecs across, in which
          brighter region represent higher surface densities; along
          the bottom row is the corresponding source caustic in blue
          and image caustic, a.k.a the critical curve, in red, both determined for a source at redshift $z_{s}=2$}
	\label{CompareFourPhysics}
\end{figure*}
The baryonic processes as implemented in the hydrodynamical
simulations affect the details of the structure of each
cluster halo. 
Figure~\ref{CompareFourPhysics} shows a single cluster of mass $\mvir=8\times10^{14}\,h^{-1}\msol$ in our $z_{L}=0.5$
sample as found in each of the four re-simulations. The basic
structure is recognisable in each simulation and the subhaloes appear
in much the same positions, as seen in the convergence maps along the
top row. However, the distribution of matter near the clusters centre
is sufficiently altered by baryonic physics so that the caustics, as
shown in the bottom row of the same figure, are distinct. In
particular, note that the {\tt CSF} simulation produces larger source
caustics, shown in blue, and larger tangential image caustics --- or critical
curves --- shown in red. The tangential part of the image caustic
extends around a substructure that was sub-critical in the other
simulations. For this particular cluster, as seen from one line of
sight, the strong lensing efficiency is highest when the cluster is
generated within the {\tt CSF} simulation.

It is our aim to determine whether this is generally true for
simulated galaxy clusters, and in particular, how the strong lensing
properties predicted by the {\tt AGN} simulations compare with the
other simulations. We have a large sample of galaxy cluster, but we
should not consider only one line of sight through each cluster.
%
\begin{figure}
      	\includegraphics[trim=2mm 0mm 0mm 10mm, clip,
        width=0.99\linewidth]{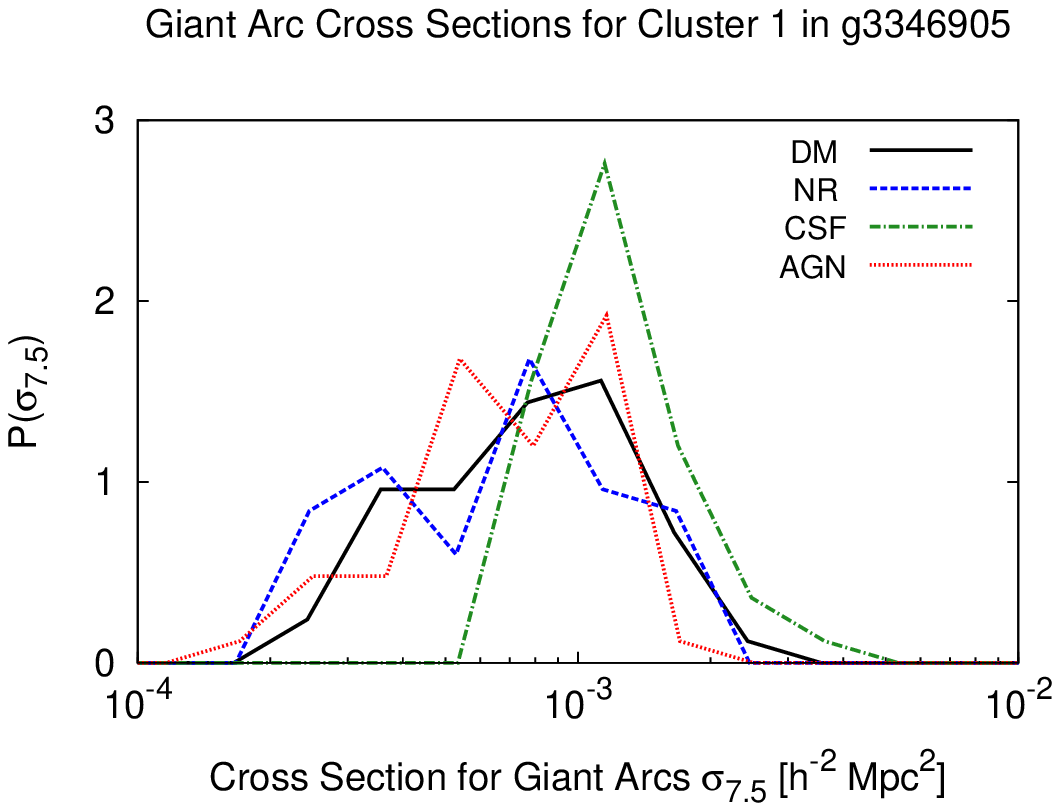}
      	\caption{The probability distributions for the cross section
          that would be measured for a single simulated cluster at
          $z_{L}=0.5$ for a source redshift of $z_{s}=2$; the cluster is the same as that shown in Figure~\ref{CompareFourPhysics}. Different
          curves refer to the distributions produced by the four different
          simulations: {\tt DM} (black solid line), {\tt NR} (blue
          dashed line), {\tt CSF} (green dash-dotted line), {\tt AGN}
          (red dotted line)}
	\label{CShistoOneCluster}
\end{figure}
Due to the aspherical nature of clusters, as well as the presence of
substructure, each line of sight produces a unique cross
section. Therefore, each cluster is associated with a distribution of
possible cross sections \citep[see, for example, figure 3
of][]{DHH04}. In Figure~\ref{CShistoOneCluster}, we show such a
distribution by measuring the cross section, $\sigma_{7.5}$, for 50
lines of sight through one cluster in our sample, also seen in Figure~\ref{CompareFourPhysics}.
%
\begin{figure*}
      \begin{minipage}{0.495\textwidth}
	\begin{flushleft}
	 \includegraphics[width=0.99\linewidth]{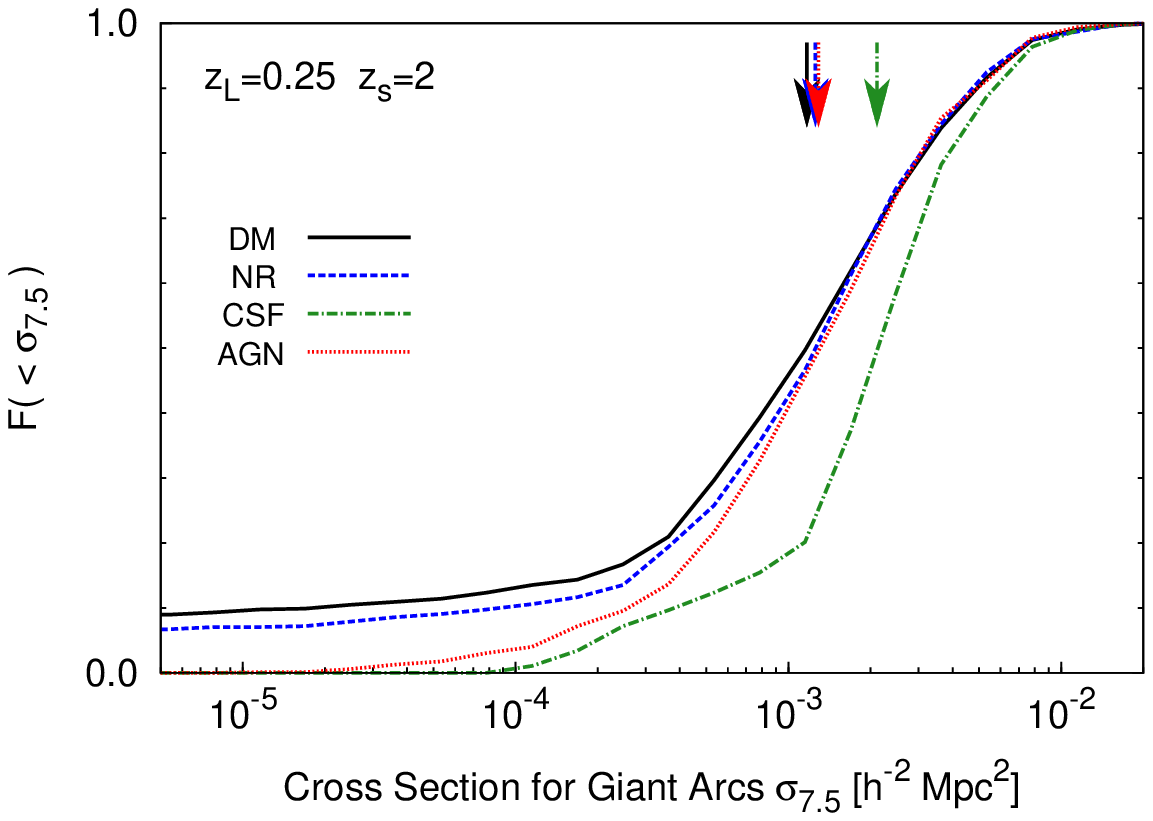}
	\end{flushleft}
      \end{minipage}
      \hfill
      \begin{minipage}{0.495\textwidth}
	\begin{flushright}
      	\includegraphics[width=0.99\linewidth]{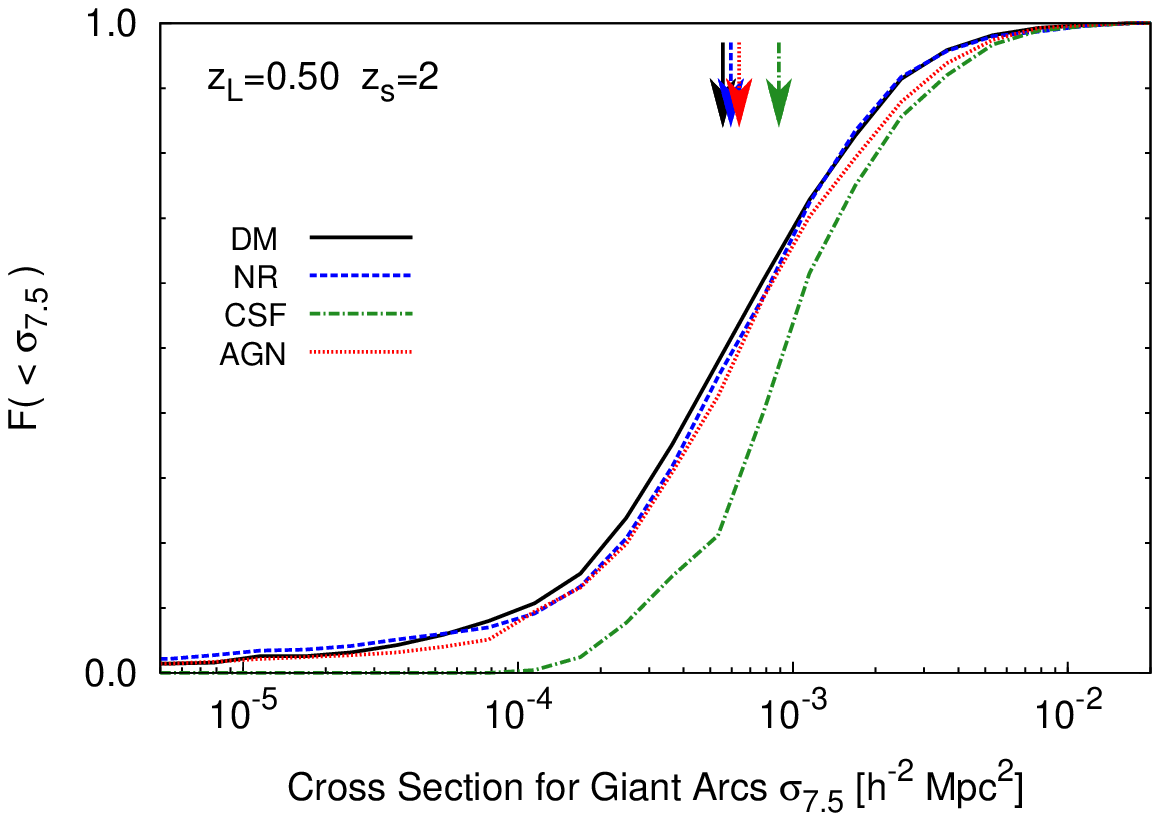}
	\end{flushright}
      	\end{minipage}
      	\caption{The cumulative probability distribution for the
          source-plane cross-section for the formation of giant arcs
          combining 50 lines of sight through the relaxed clusters at
          $z_{L}=0.25$ (left panel) and $z_{L}=0.5$ (right
          panel). Different lines and colours have the same meaning as
          in Figure \protect\ref{CShistoOneCluster}. The arrows mark the
          median values of the $\sigma_{7.5}$ cross section for the
          models. }
	\label{CrossSecPDF}
\end{figure*}
In the present study, we analyse projections of each cluster along 50
randomly chosen lines of sight. The results for all our sets of
simulated clusters are combined in the
left-hand panel of Figure~\ref{CrossSecPDF}.

We conduct a two-sample Kolmogorov-Smirnov test to compare the probability
distribution of $\sigma_{7.5}$ values for each simulation against the
same distribution for the {\tt DM} simulations. The D-statistic is the maximum difference
between the two cumulative probability distributions. Ideally we would like to determine the probability,
$p_{\sigma}$, that the D-statistic would be atleast as large as that measured assuming the samples are drawn from the same underlying distribution. We recognise that while the clusters are independent, the
many lines of sight analysed for each cluster are not independent from
each other. In an actual observational sample, one would measure a
single strong-lensing efficiency for each cluster; however using a
simulated sample, scatter due to orientation has to be taken into
account. Therefore, stacking the results for all lines of sight
analysed, we calculate the D-statistic. Using the total number of lines of sight (50 times the number of clusters) as the `sample size' provides a lower limit for the p-value,
$p_{\sigma}^{min}$. Alternatively, by letting the number of clusters
be the `sample size', we can obtain an upper limit on the p-value,
$p_{\sigma}^{max}$. Both are listed in Table 1.  If the clusters were all perfectly spherically symmetrical, so that all lines of sight resulted in the same value of $\sigma_{7.5}$, then $p_{\sigma}^{max}$ would be equivalent to the required $p_{\sigma}$.

Table 1 also include the typical cross-sections for
clusters in each simulation, for low and high source redshifts:
$z_{s}=1$ and $z_{s}=2$. These are calculated by taking the median
value of the cross-section over the 50 lines of sight analysed for
each clusters, then averaging over all clusters. Comparing the results
for {\tt DM} and {\tt NR} simulations, we find that the ability to
lens background galaxies into giant arcs is similar, independently of
whether the cluster lens is simulated with collision-less particles
only, or with additional non-radiative gas.
As seen for the {\tt CSF} simulation, when cooling and star-formation
are implemented, the cross-section for the formation of giant arcs is
boosted by a factor of $\sim1.5$ for high source redshifts ($z_{s}=2$)
and $\sim2$ for lower source redshifts ($z_{s}=1$). This is a slightly
smaller `boost' relative to the findings of \citet{P05}, \citet{R08}
and \citet{Mead10}.
This difference with respect to previous simulation models can be
understood in terms of the more efficient SN feedback implemented in
our CSF simulation set, which more efficiently counteracts the effect
of halo contraction induced by cooling. In fact, \citet{P05} use the same model for galactic ejecta, but with a smaller velocity of galactic winds, with $v_w=350$ km s$^{-1}$. Furthermore, \citet{R08} and \citet{Mead10} only include thermal schemes for SN feedback, which are rather inefficient in regulating cooling at the centre of cluster--sized halos.

AGN feedback reduces the giant-arc cross-section and makes it
comparable to the predicted cross-section from collision-less
simulations. When we compare clusters modelled with dark matter only
({\tt DM} results shown in black) and their counterparts modelled with
the complete set of baryonic processes, including AGN feedback ({\tt
  AGN} shown in red), we find surprisingly little difference in the
typical cross-sections for clusters at $z_{L}=0.25$ - now a boost of
only $\sim10$ per cent. However, at $z_{L}=0.5$ clusters modelled with AGN
feedback are $\sim30$ per cent more efficient at lensing background
galaxies into giant arcs. Given the results for Einstein radii (see
Section~\ref{baryonsER} below), this is likely due to the presence of
substructures that produce additional arcs rather than the
re-distribution of mass near the cluster centre.  The difference in the results between {\tt CSF} and the {\tt AGN} simulations arise from a combination of the reduced SN wind-speed and the introduction of thermal AGN feedback. However, if the {\tt CSF} simulations had a similarly reduced wind-speed, the reduced feedback efficiency would lead to increased cooling and a higher lensing cross-section, therefore it is the AGN feedback that is responsible for reducing the lensing efficiency.
\citet{Mead10} analysed three projections for each of five clusters at $z_{L}=0.2$
with similar simulation sets. For four of these five clusters the
strong lensing cross-sections for the dark matter simulation and their
AGN simulation were very similar. At this redshift, given the small
sample and number of projections, this is consistent with our
findings.

%
\begin{table*}
\label{t:statsTable}
\begin{minipage}{140mm}
  \caption{Statistical Analysis of Strong Lensing Properties of
    Relaxed Clusters. Column 1: simulation set. Columns 2 and 3:
    redshifts of the lens, $z_L$, and of the source, $z_s$. Column 4:
    Pearson correlation coefficient $r$ for the
    log($\sigma_{7.5}$)--log($\EinRad$) relation. Column 5 and 6: least-squares fit to the log($\sigma_{7.5}$)--log($\EinRad$) relation (see Eqn.~\ref{CSERrelation}). Column 7: typical
    value of the Einstein radius, $\EinRad$, computed as the median
    value over 50 lines of sight through each cluster, then averaged
    over all clusters (units of arcseconds). Columns 8 and 9: values
    of $p_{\theta}^{min}$ and $p_{\theta}^{max}$, defined as the lower
    and upper limits, respectively, of a true $p$-value from a KS-test
    performed to compare the $\EinRad$ probability distribution
    of each simulation set against the corresponding result for the
    {\tt DM} simulation. Column 10: the typical cross section for the
    lensing of galaxies at $z_{s}$ into giant arcs with elongation
    $\eta=7.5$, $\sigma_{7.5}$, defined as the median value over 50
    lines of sight through each cluster, then averaged over all
    clusters (units of $10^{-4}h^{-2}$Mpc$^{2}$). Columns 11 and 12:
    the same as in columns 8 and 9, respectively, but to compare the
    $\sigma_{7.5}$ probability distribution.
    }
\begin{center}
\begin{tabular}{@{\extracolsep{\fill}} l c l c c c c c c c c c} 
\hline
Simulation & $z_L$ & $z_s$ 	& r & a & b & $\EinRad$ & $p_{\theta}^{min}$ & $p_{\theta}^{max}$ & $\sigma_{7.5}$  & $p_{\sigma}^{min}$ & $p_{\sigma}^{max}$ \\ [1ex]
\hline \hline
DM 		& 0.25 & 2  & 0.94 	& 2.15 $\pm$ 0.02 & -6.04 $\pm$ 0.03 & 29 	 & -- 				& -- 	 	& 16  	& -- 				& -- 	 \\
NR 		& 0.25 & 2  & 0.92 	& 1.93 $\pm$ 0.03 & -5.70 $\pm$ 0.04 & 30 	 & 0.02 			& 1.00  	& 17  	& 0.11 			& 1.00  \\
CSF		& 0.25 & 2  & 0.97 	& 2.02 $\pm$ 0.02 & -5.72 $\pm$ 0.03 & 33 	 & $<10^{-5}$  		& 0.87 	& 24  	& $<10^{-5}$  		& 0.35 \\
AGN		& 0.25 & 2  & 0.97 	& 2.11 $\pm$ 0.02 & -5.96 $\pm$ 0.03 & 30 	 & $1\times10^{-4}$	& 1.00	& 18  	& $3\times10^{-4}$  	& 1.00 \\
\hline
DM 		& 0.5 & 2  & 0.96  	& 2.10 $\pm$ 0.02 & -5.90 $\pm$ 0.03 & 20 	& -- 				& -- 		& 8.2  	& -- 				& -- 	\\
NR 		& 0.5 & 2  & 0.95 	& 2.08 $\pm$ 0.02 & -5.89 $\pm$ 0.03 & 21 	& 0.11			& 1.00	& 8.9  	& 0.14 			& 1.00 \\
CSF		& 0.5 & 2  & 0.96  	& 1.87 $\pm$ 0.02 & -5.45 $\pm$ 0.03 & 21 	& $7\times10^{-5}$	& 1.00	& 12.4  	& $<10^{-5}$		& 0.59 \\
AGN		& 0.5 & 2  & 0.96  	& 2.09 $\pm$ 0.02 & -5.84 $\pm$ 0.03 & 21 	& 0.93 			& 1.00	& 10.6  	& 0.07 			& 1.00 \\
\hline
DM 		& 0.25 & 1  & 0.94  	& 2.02 $\pm$ 0.02 & -5.92 $\pm$ 0.03 & 21	& -- 				& -- 	 	& 5.0  	& -- 				& -- 	 \\
NR 		& 0.25 & 1  & 0.91 	& 1.81 $\pm$ 0.03 & -5.65 $\pm$ 0.04 & 22	& 0.008 			& 1.00  	& 5.6  	& 0.16 			& 1.00 \\
CSF		& 0.25 & 1  & 0.96  	& 2.10 $\pm$ 0.02 & -5.92 $\pm$ 0.03 & 25	& $<10^{-5}$  		& 0.80 	& 10.5  	& $<10^{-5}$  		& 0.11 \\
AGN		& 0.25 & 1  & 0.95  	& 2.18 $\pm$ 0.02 & -6.14 $\pm$ 0.03 & 22 	& $<10^{-5}$ 		& 0.99	& 6.1  	& $<10^{-5}$ 		& 1.00 \\
\hline
DM 		& 0.5 & 1 	& 0.89 	& 2.21 $\pm$ 0.04 & -6.13 $\pm$ 0.04 & 9 	& -- 				& -- 	 	& 1.6  	& -- 				& -- \\
NR 		& 0.5 & 1  	& 0.88 	& 2.29 $\pm$ 0.04 & -6.29 $\pm$ 0.05 & 10 & $4\times10^{-4}$	& 1.00  	& 1.5  	& 0.20 			& 1.00  \\
CSF		& 0.5 & 1 	& 0.93 	& 2.02 $\pm$ 0.03 & -5.78 $\pm$ 0.03 & 12 	& $<10^{-5}$	 	& 0.46 	& 2.9  	& $<10^{-5}$  		& 0.25 \\
AGN		& 0.5 & 1	& 0.92  	& 2.32 $\pm$ 0.03 & -6.22 $\pm$ 0.04 & 10 	& $<10^{-5}$ 		& 0.95	& 2.2  	& $<10^{-5}$ 		& 0.99 \\
\hline
\end{tabular}
\\
\end{center}

\end{minipage}
\end{table*}
%

%
\begin{figure*}
      \begin{minipage}{0.495\textwidth}
      \begin{flushleft}
      \includegraphics[width=0.99\linewidth]{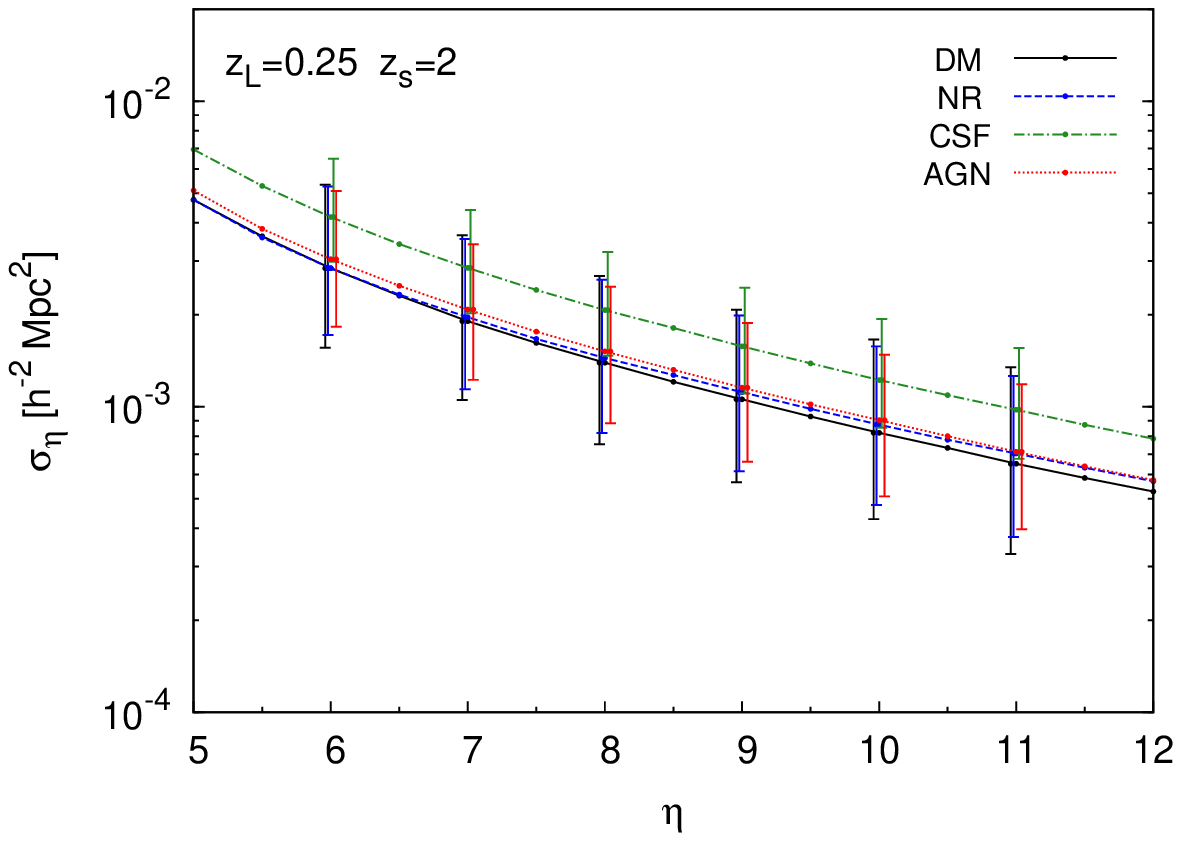}
      \end{flushleft}
      \end{minipage}
      \hfill
      \begin{minipage}{0.495\textwidth}
      \begin{flushright}
      \includegraphics[width=0.99\linewidth]{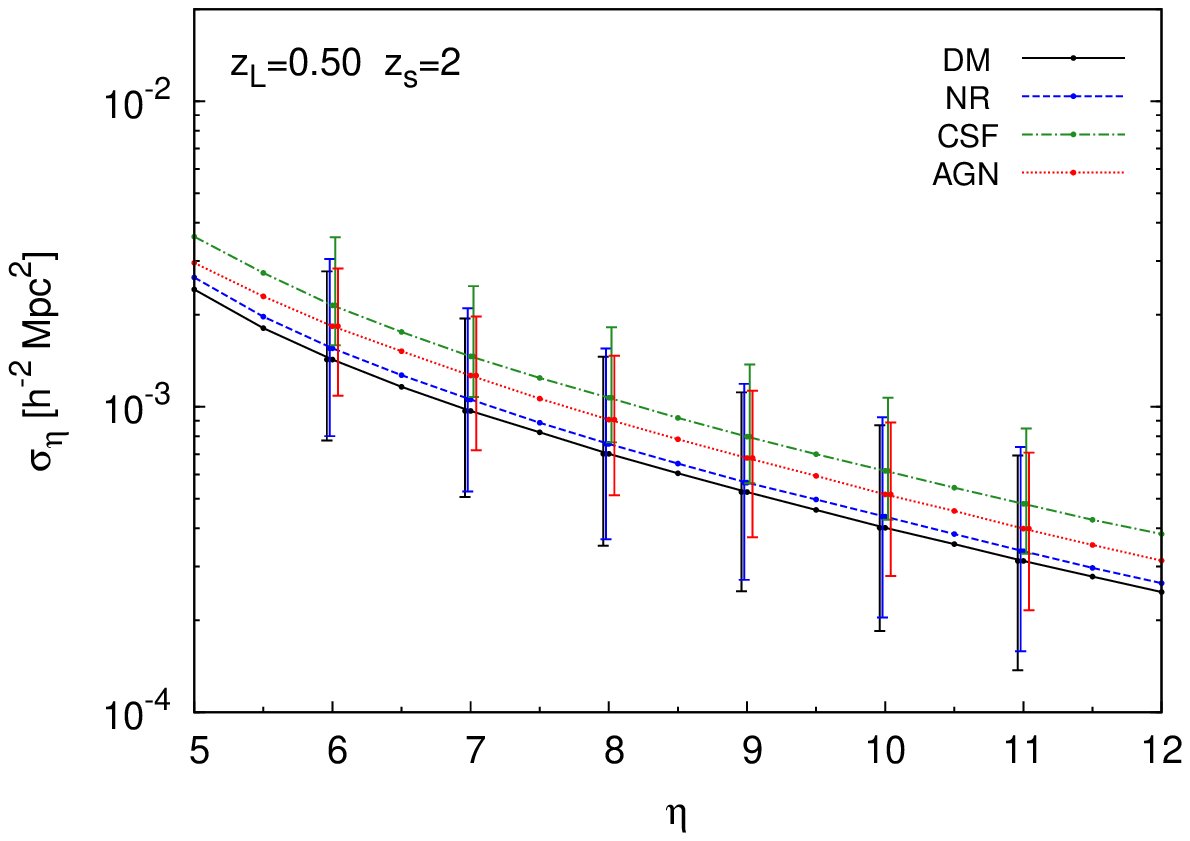}
      \end{flushright}
      \end{minipage}
      \caption{The giant arc cross section plotted against chosen
        elongation threshold $\eta$ for the relaxed cluster sub-sample at
        $z_{L}=0.25$ (left panels) and $z_{L}=0.5$ (right panels). The
        cross section for a given value of $\eta$ is the median of 50
        lines of sight through each cluster, then averaged over all
        clusters. The errorbars mark the 16th and 84th percentile for
        each cluster, then averaged over all clusters. The source
        redshift is $z_{s}=2$}
       \label{CrossSecEta}
\end{figure*}
As described earlier, images are identified as arcs if their
length-to-width ratio surpasses the threshold elongation,
$\eta$. Although extreme values for this threshold are not useful, the
exact value is somewhat arbitrarily chosen; to be sure that the
results are not qualitatively affected by this choice, we plot the
cross section as a function of $\eta$ in Figure~\ref{CrossSecEta}. The
lines represent the median cross section over the 50 lines of sight
for each individual cluster, then averaged over all the clusters in
the (relaxed) sample. The error bars represent the 16th and 84th
percentile for individual clusters, then averaged over all the
clusters. These error-bars, therefore, reflect the triaxiality of the
clusters.
Unsurprisingly, a larger and more extreme choice of threshold allows a
smaller cross section, since less sources would be able to produce
such long and thin arcs, but the qualitative differences between the
different simulation sets are the same.

By comparing the error-bars in Figure~\ref{CrossSecEta} for all
re-simulations, we find that as in \citet{R08}, the spread in the
distribution of possible cross-sections is also reduced significantly
in the {\tt CSF} simulations.  These clusters are less susceptible to
light-of-sight effects than those modelled with dark matter only or
non-radiative gas, predominantly due to the more spherical shape of
the clusters resulting from the condensing of baryons and the
subsequent response from the dark matter component \citep{BFFP86,G04,K04,Bryan12}. The steeper inner profile (see
Section \ref{profiles}) makes the clusters less susceptible to
coincidental mass along the line of sight. Although AGN feedback
regulates the isotropic condensation of baryons, the reduced spread in
cross-sections is also found for the {\tt AGN} clusters, particularly
at $z_{L}=0.5$.

\begin{figure*}
      \begin{minipage}{0.49\textwidth}
	\begin{flushleft}
	 \includegraphics[width=0.99\linewidth]{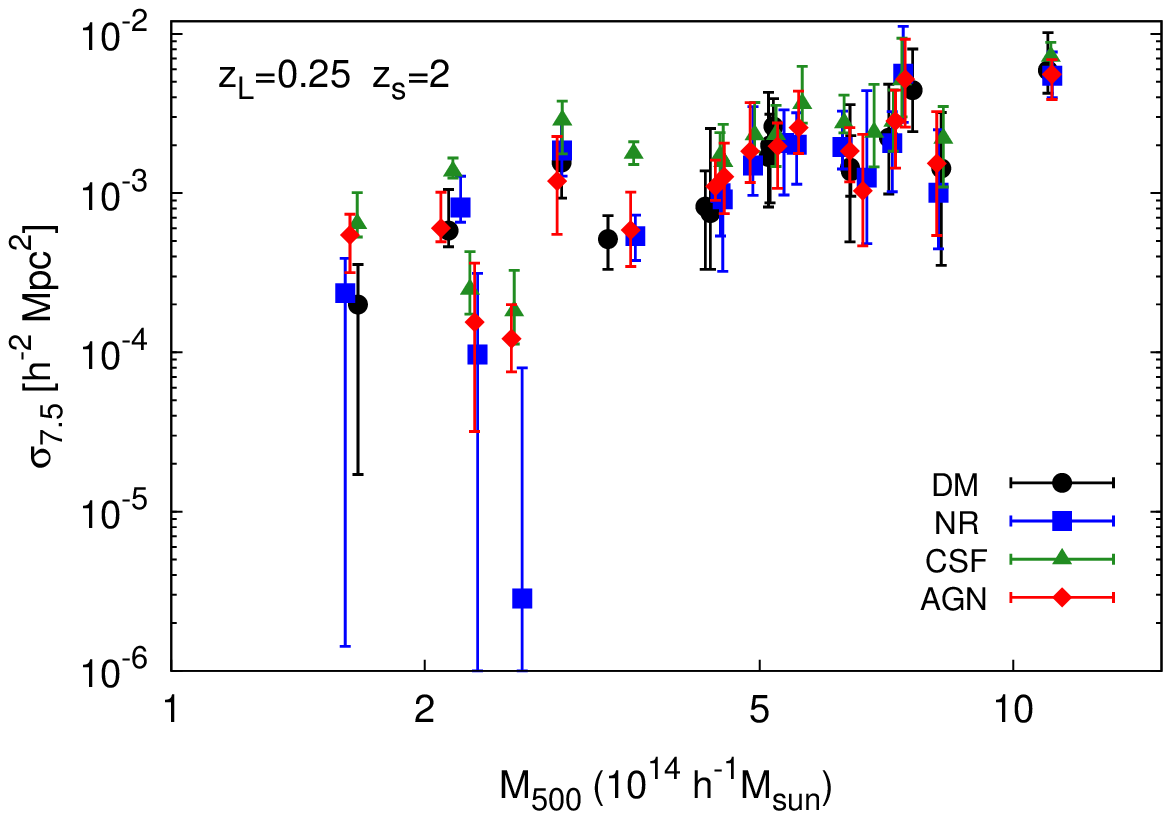}
	 \includegraphics[width=0.99\linewidth]{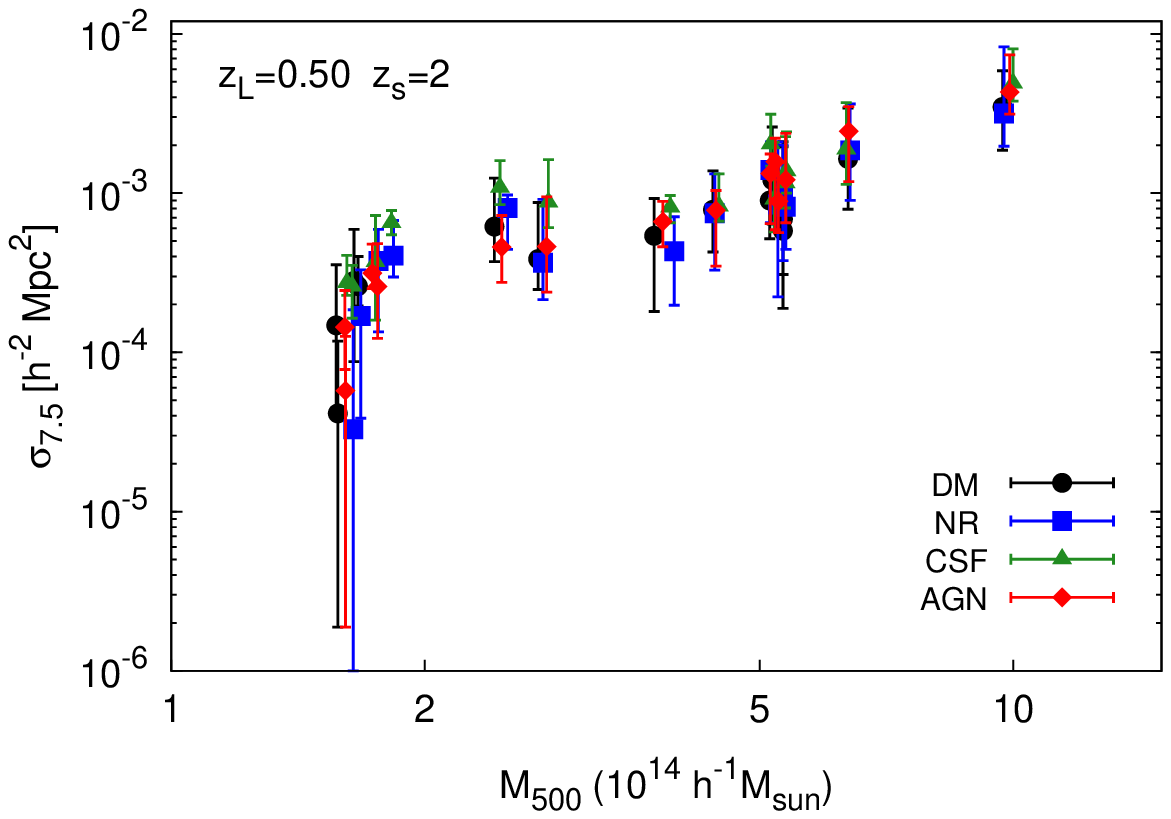}
	\end{flushleft}
      \end{minipage}
      \hfill
      \begin{minipage}{0.49\textwidth}
	\begin{flushright}
	 \includegraphics[width=0.99\linewidth]{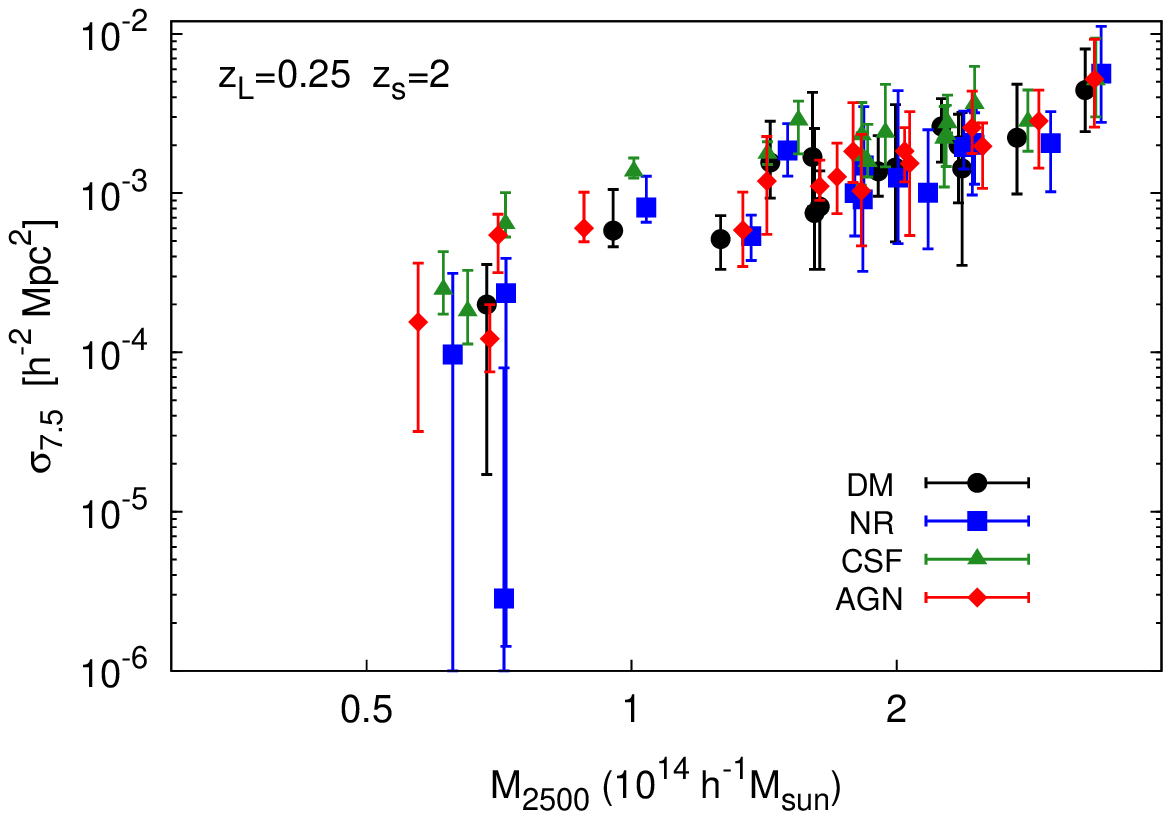}
	 \includegraphics[width=0.99\linewidth]{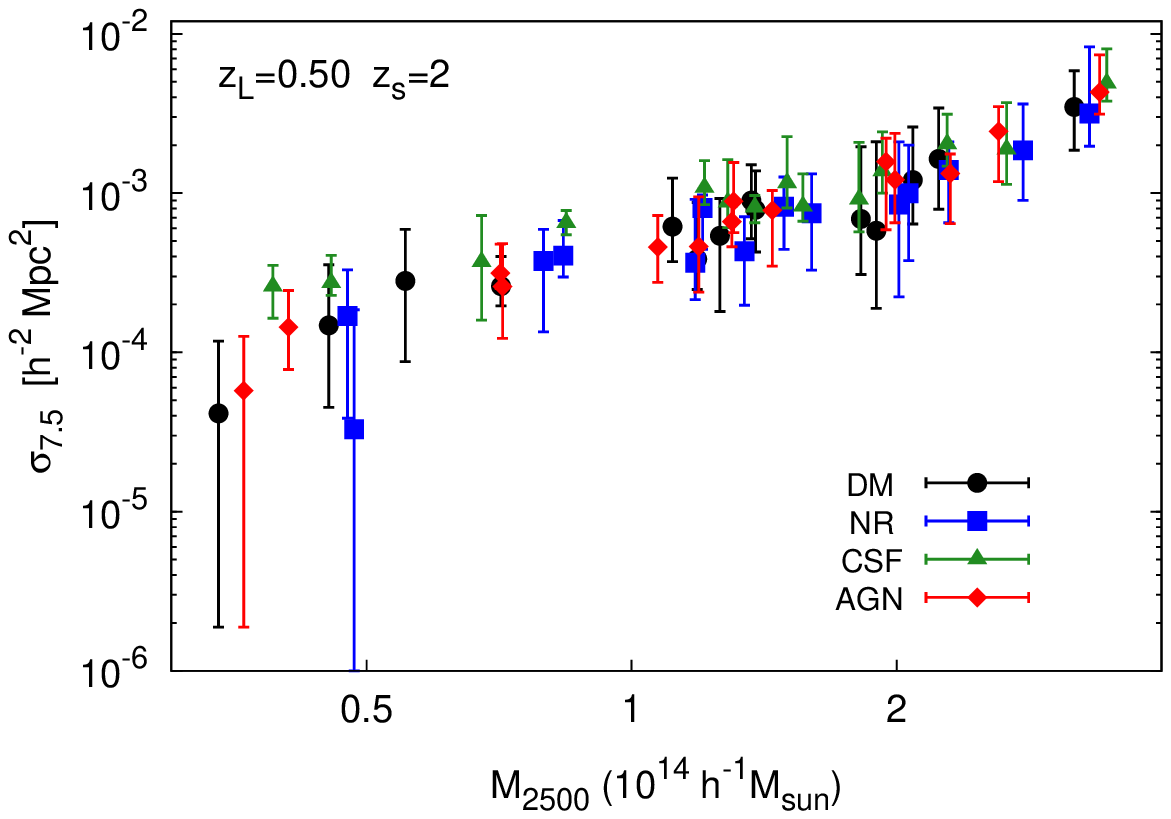}
	\end{flushright}
      	\end{minipage}
      	\caption{The cross-section for the formation of giant arcs at
          source redshift of $z_{s}=2$ as a function of cluster mass,
          for our relaxed cluster sub-sample. Each dot represents the
          median cross-section from the 50 lines of sight analysed for
          a single cluster, while the error bars mark the 16th and
          84th percentiles. In the left-hand panels, the
          characteristic mass is $M_{500}$ while for the right-hand
          panels, the characteristic mass is $M_{2500}$. The clusters
          at $z_{L}=0.25$ are shown on the top row while the clusters
          at $z_{L}=0.5$ are shown on the bottom row.}
	\label{csVmass}
\end{figure*}
%
Figure~\ref{csVmass} shows the giant arc cross-section, $\sigma_{7.5}$,
plotted against the mass of each clusters. Each cluster is represented
as a single point in the graph, with different colours representing
the counterparts in the four re-simulations. The error-bars mark the
16th and 84th percentiles of $\sigma_{7.5}$ found over the 50 lines of
sight analysed. The figure suggests that there exists a scaling
relation between lensing efficiency and mass, but that the correlation
is tighter at higher-overdensities; this reflects the region most
responsible for strong lensing. The higher scatter when $\sigma_{7.5}$
is plotted as a function of M$_{500}$ reflects the fact that different
re-simulations create differing distributions of baryons in the
cluster core, when comparing clusters at a fixed mass at lower
overdensities.

\section{Einstein Radii}\label{ER} 
The cosmological test based on arc statistics is subject to
uncertainty in the characteristics of the source
population. Performing a comparison with observed arc statistics
requires one to convolve the predicted giant arc cross-section with an
assumed redshift distribution for the background galaxies. The
uncertainties in this redshift distribution creates an additional
uncertainty in the cosmological test. This additional uncertainty is
unnecessary since one might instead characterise strong lensing
efficiency by the angular scale which separates highly magnified
images; this is known as the Einstein radius. Measuring the Einstein
radii requires the observer to: measure the shape and redshift of a
reasonable number of highly magnified high redshift sources,
reconstruct the lens mass distribution, and infer the critical curves
at a single redshift. This is achievable even if the various sources
are at different redshifts, and bypasses the need to independently
measure the expected number density and redshift evolution of the
sources.
We therefore measure the Einstein radii for our cluster sample and
compare the results between the different re-simulations.

\subsection{Measuring the Einstein Radius}\label{findER}
The angular separation of highly magnified background galaxies has a
formal definition, which is strictly applicable only in the case of
axially symmetric lenses. However, galaxy clusters are not axially
symmetric in general; there exist a number of different definitions in
the literature for how one might calculate the Einstein radius for
more realistic lenses. For example, one might divide the area enclosed
within the tangential critical curve by $\pi$, as per the so-called
`equivalent Einstein radius' definition used, e.g., by \citet{PH09}, \citet{Z11a}
and \citet{R12}. On the other hand, \citet{BB08} define the `effective
Einstein radius' as the radius that encloses a mean surface density
equal to the critical surface density $\kappa=1$ (see \citealt{PH09} for further discussion on how these two definitions compare).
We, instead, follow the definition of \citet{M11} and characterise our statistics
by a `median Einstein radius': the median distance of the tangential
critical points from the clusters centre. \citet{M11} has demonstrated
that the median Einstein radius has a tighter correlation with strong
lensing cross section than the `equivalent Einstein radius'. When
  measuring the Einstein radius, \citet{M11} define the cluster centre
  to be the location of the maximum of the projected mass
  distribution. In the present work, we define the centre to be the
  projected position of the particle in the simulated cluster with the
  lowest potential. This is done in order to avoid attributing the
  position of the centre to a subhalo in cases in which such a subhalo
  is fortuitously projected so it has a higher projected mass density
  than the centre of the main halo.

We use the high resolution deflection-angle map (as described in
Section~\ref{findXsec}) to identify tangential critical points within
the same field of view at the same angular resolution. There can be
complications in the presence of substructure, which require one to
discard some of the critical points before measuring the Einstein
radius.
%
\begin{figure*}
      \begin{minipage}{0.47\textwidth}
      \begin{flushleft}
      \raisebox{8mm}{\includegraphics[trim=0mm 50mm 0mm 50mm, clip, width=\linewidth]{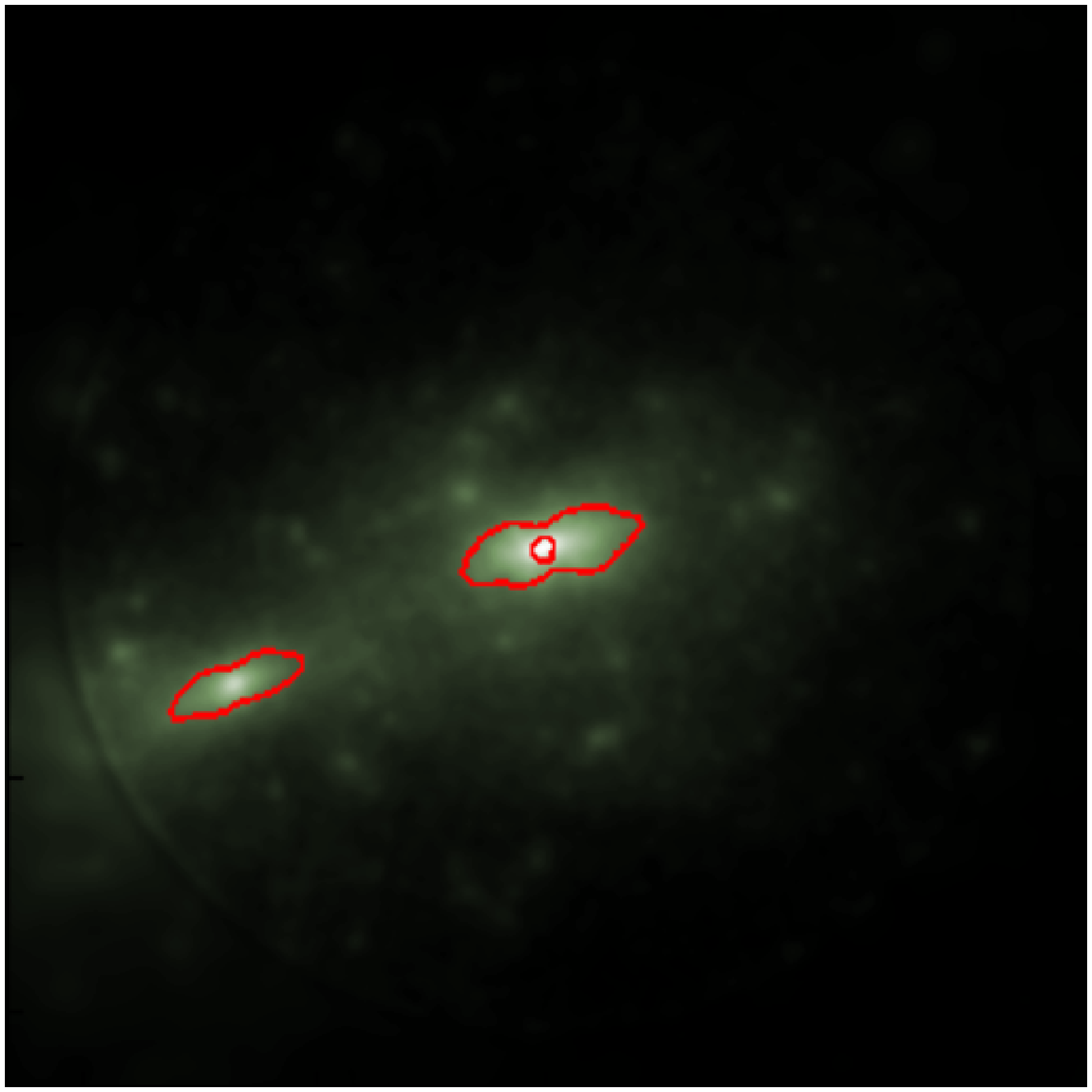}}
      \includegraphics[trim=0mm 50mm 0mm 50mm, clip, width=\linewidth]{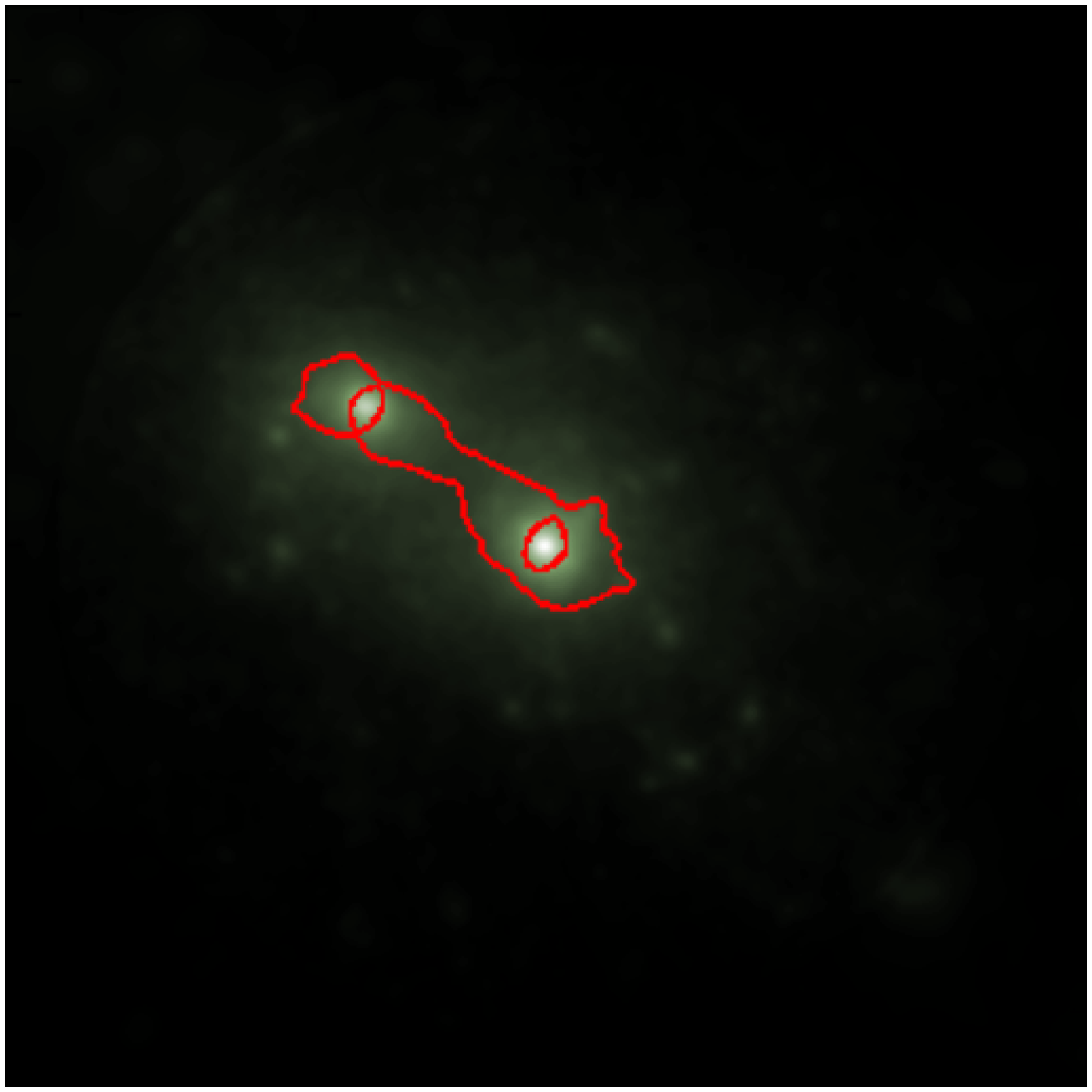}
      \end{flushleft}
      \end{minipage}
      \hfill
      \begin{minipage}{0.52\textwidth}
      \begin{flushright}
      \includegraphics[trim=0mm 15mm 0mm 10mm, clip, width=\linewidth]{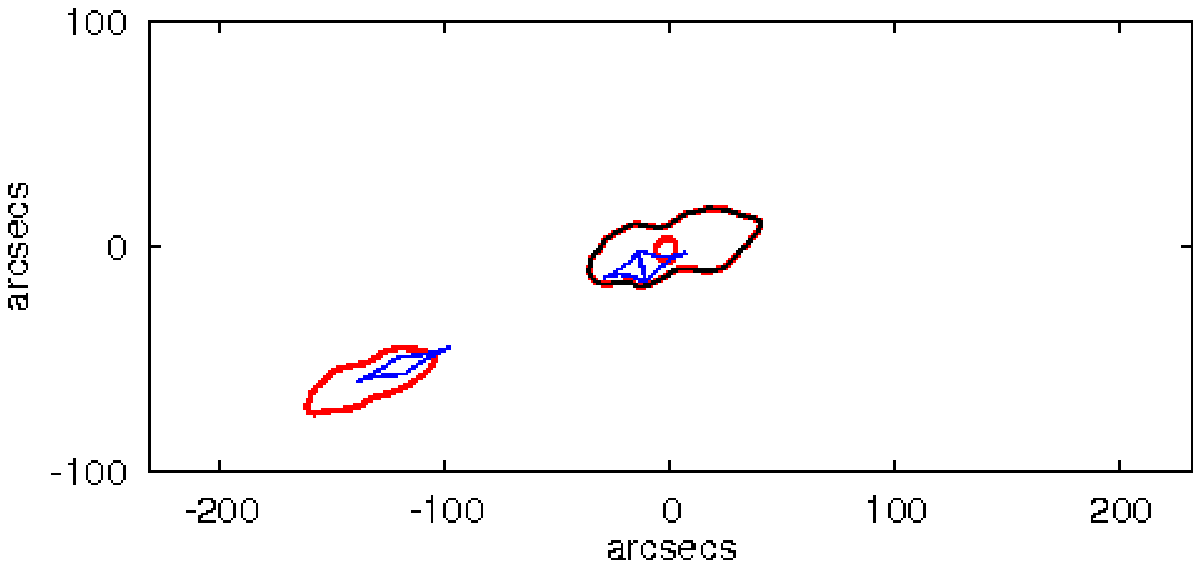}
      \includegraphics[trim=0mm 15mm 0mm 10mm, clip, width=\linewidth]{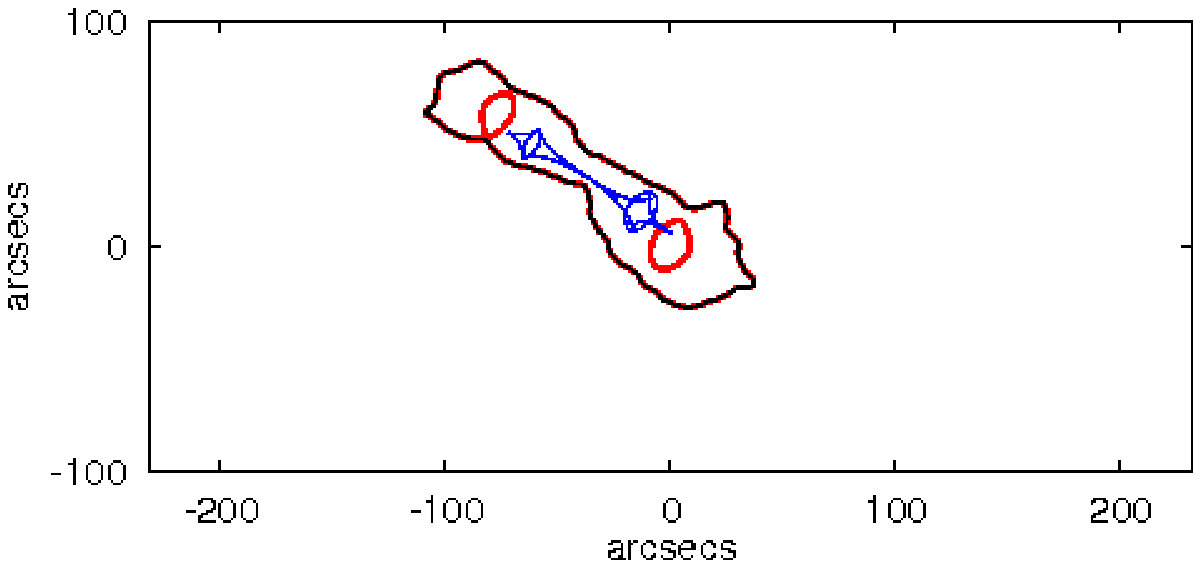}
      \end{flushright}
      \end{minipage}
      \caption{The surface density map (left-hand panels) and image
        and source caustic (right-hand panels) of an {\it unrelaxed}
        $z_{L}=0.5$ cluster in our sample. In the left-hand panels,
        brighter regions correspond to higher surface density and the
        image caustic is overlaid in red. In the right-hand panels,
        the image caustic is shown in red, with the `cleaned' caustic
        - used to infer the Einstein radius - drawn in black, and the
        source caustic at $z_{s}=2$ shown in blue. The `primary' image
        caustic is associated with the centre of the cluster, while a
        large substructure present within the field of view produces a
        `secondary' image caustic. Seen from one line of sight (top
        row), the secondary caustic is distinct, so the cleaned
        caustic captures only the primary caustic, while from the
        other line of sight (bottom row) the primary and secondary
        caustics are merged, so the cleaned caustic captures both, and
        artificially enhances the inferred Einstein radius.}
       \label{SubsCaus}
\end{figure*}
%
We show an unrelaxed cluster in Figure~\ref{SubsCaus} as an example. There is a large substructure present and, from some lines of sight, visible within the field of view. For each line of sight analyzed, critical points are mapped out across the lens plane forming a `critical curve' also known as an image caustic; these correspond to the caustics in red. Then, critical points associated with the tangentially sheared images (as opposed to radially sheared) at the `primary' critical curve around the centre of the cluster are identified, so that `secondary' critical curves associated with substructure do not bias the measurement. These are the `cleaned' curves shown in black. The projected distance of each `clean' point from the cluster centre is measured; the Einstein radius, $\EinRad$, is the median value over all these points.

 If the secondary critical curve merges with the primary, as shown in
 the bottom row of Figure~\ref{SubsCaus}, they cannot be
 distinguished, and $\EinRad$ is artificially increased. The
 cross-section $\sigma_{7.5}$ gives, to a first approximation, the
 region enclosed by
 the source caustics (shown in blue). The measurement of $\sigma_{7.5}$
 can be artificially enhanced by the presence of substructure. We
 discuss this further in Section \ref{unrelaxed}. For a relaxed cluster, there would be fewer lines of sight in which the secondary and primary critical curves merge, and the cluster will have smaller secondary critical curves which have a less significant impact on the measurement of $\sigma_{7.5}$ and $\EinRad$.

%
\begin{figure*}
	\begin{minipage}{0.495\textwidth}
	\begin{flushleft}
      	\includegraphics[width=0.99\linewidth]{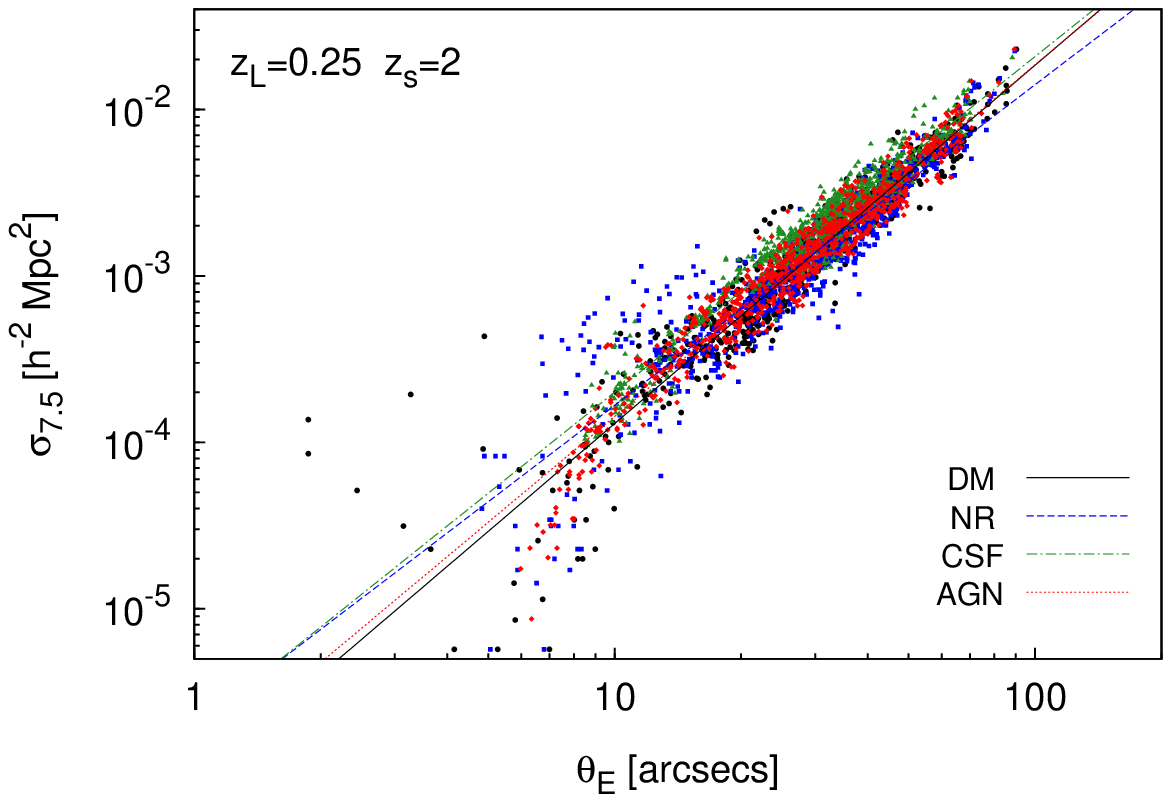}
	\end{flushleft}
	\end{minipage}
	\hfill
	\begin{minipage}{0.495\textwidth}
	\begin{flushright}
      	\includegraphics[width=0.99\linewidth]{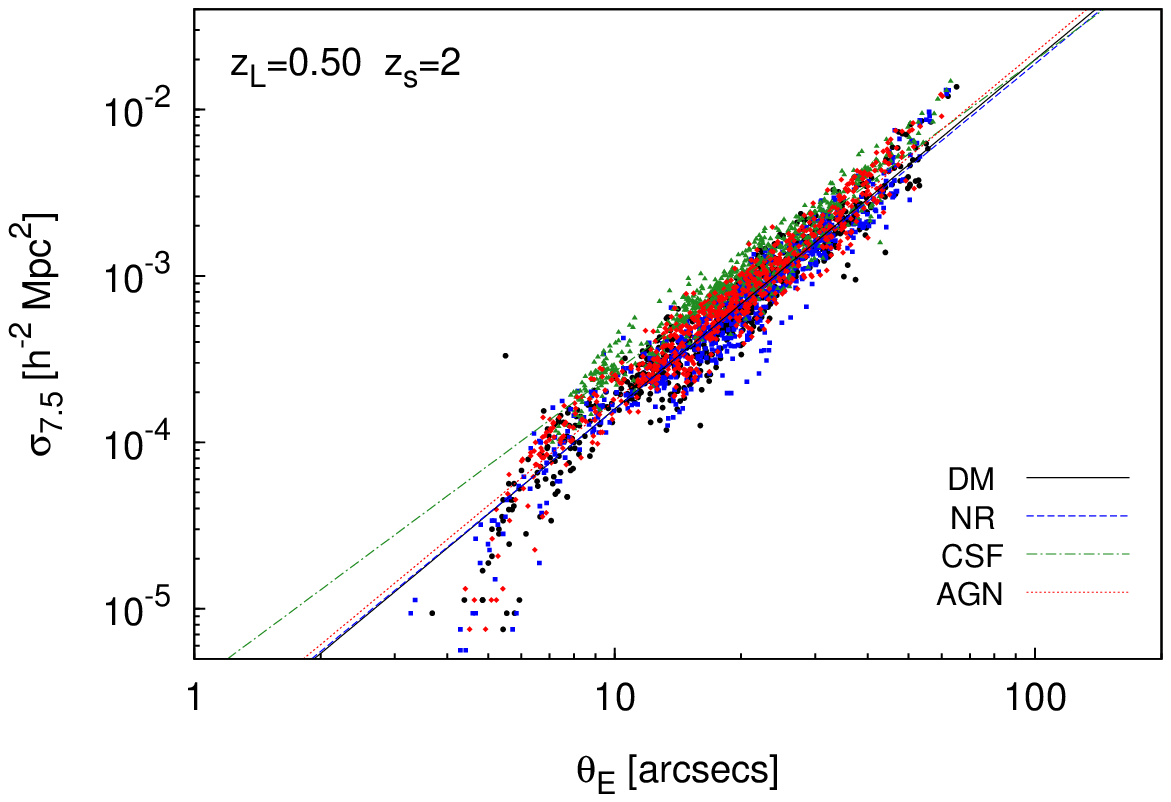}
	\end{flushright}
	\end{minipage}
      	\caption{Relationship between the giant-arc cross section,
          $\sigma_{7.5}$, and the Einstein radius, $\EinRad$, for each
          of the 50 orientations of each cluster. On the left panel,
          we show the results from the clusters at $z_{L}=0.25$ and on
          the right panel, we show the results from the clusters at
          $z_{L}=0.5$; both are for source redshift of
          $z_{s}=2$. Results for the four simulations are combined:
          dark matter only (black); dark matter and non-radiative gas
          (blue); with cooling and star-formation (green); and with
          AGN feedback (red)}
	\label{LensStatsCorrelate}
\end{figure*}
Figure~\ref{LensStatsCorrelate} shows the giant-arc cross section and the Einstein radii for corresponding lines of sight through each cluster. As in \citet{M11}, it is clear that the correlation is very strong for our relaxed cluster subsample. We perform a least-squares fitting to a function of the form:
\begin{equation}\label{CSERrelation}
\mathrm{log}(\sigma_{7.5}) = a \mathrm{log}(\EinRad) + b,
\end{equation}
and measure the Pearson correlation coefficient 
for each simulation separately, including only lines of sight with
non-zero values for both the cross section and Einstein radii. The
results for $a$, $b$ and $r$ are summarised in
Table 1. The correlation coefficient, $r$, ranges from
0.92 to 0.97 for the relaxed clusters at $z_{L}=0.25$, and from
0.95 to 0.96 for the relaxed clusters at $z_{L}=0.5$. 
Two of the clusters in the {\tt NR} simulation are responsible for highly elliptical critical curves and elongated source caustics, which results in the scatter above the line of best-fit in the left panel of Figure~\ref{LensStatsCorrelate}, and makes the line of best-fit shallower; this is indicative of the sensitivity of the fit to substructure. Similarly, a recent study by \citet{R12} has concluded that the presence of substructure and cluster mergers in the \citet{M11} simulated clusters sample --- as opposed to semi-analytic smooth triaxial cluster-halo models ---  results in shallower slope and higher normalisation. The line of best-fit to Eqn.~\ref{CSERrelation} for the {\tt NR} simulation at $z_{L}=0.5$ and $z_{s}=2$ is the most appropriate for comparison to the results of \citet{M11}, which are obtained using the $z>0.5$ clusters found in the non-radiative {\tt MareNostrum} simulation.  Compared to \citet{M11} we find a lower normalisation, $a$, and a steeper slope, $b$, of the log($\sigma_{7.5}$)--log($\EinRad$) relation. This can be attributed to our deliberate exclusion of clusters with significant substructure and merger-activity in our relaxed subsample.

We can expect the slope of the log($\sigma_{7.5}$)--log($\EinRad$) relation to reflect the inner mass profile of the lens \citep[][]{R12}. Dark-matter haloes, such as those in {\tt DM} simulations, follow NFW profiles \citep{NFW96}, with inner density profiles that fall off as $\rho(r) \propto r^{-1}$. In clusters with significant gas cooling, such as those in the {\tt CSF} simulations, the density profile is close to isothermal ($\rho \propto r^{-2}$), and so, produce a large tangential magnification relative to radial magnification, and therefore a boost in $\sigma_{7.5}$ compared to the {\tt DM} clusters. Since the positions of the critical points are relatively stable to the inner slope --- assuming no significant redistribution of mass to larger radii --- there is a greater boost in the giant arcs cross section as opposed to Einstein radius, particularly for low mass lenses \citep[see figure~5.7 in][]{O04}, which results in a shallower slope in the log($\sigma_{7.5}$)--log($\EinRad$) relation, and a smaller value of $a$. 
For low source redshifts, lensing efficiencies are lower, resulting in greater scatter as the source sizes approach the resolution of the lensing maps; the line of best-fit is sensitive to this scatter and less sensitive to the inner mass profiles.

\subsection{The impact of baryons on the Einstein radii}\label{baryonsER}
\begin{figure*}
	\begin{minipage}{0.495\textwidth}
	\begin{flushleft}
      	\includegraphics[width=0.99\linewidth]{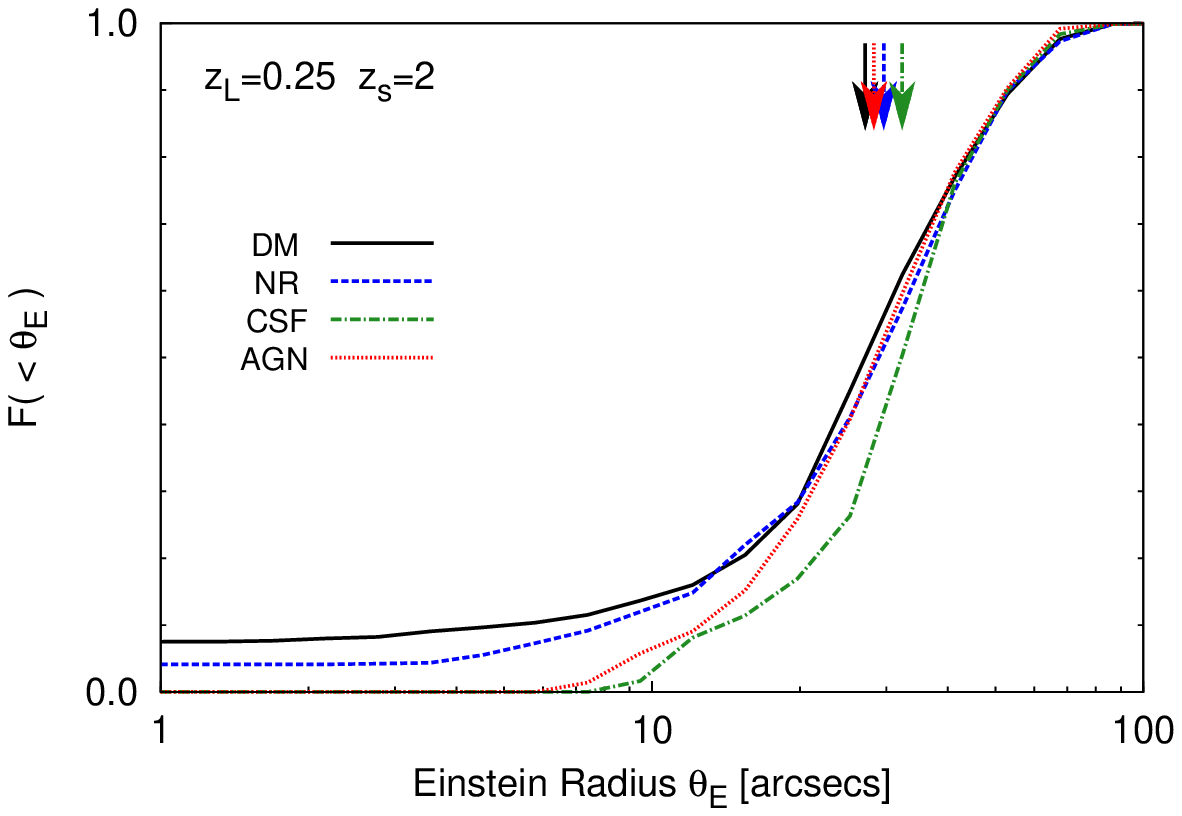}
	\end{flushleft}
	\end{minipage}
	\hfill
	\begin{minipage}{0.495\textwidth}
	\begin{flushright}
      	\includegraphics[width=0.99\linewidth]{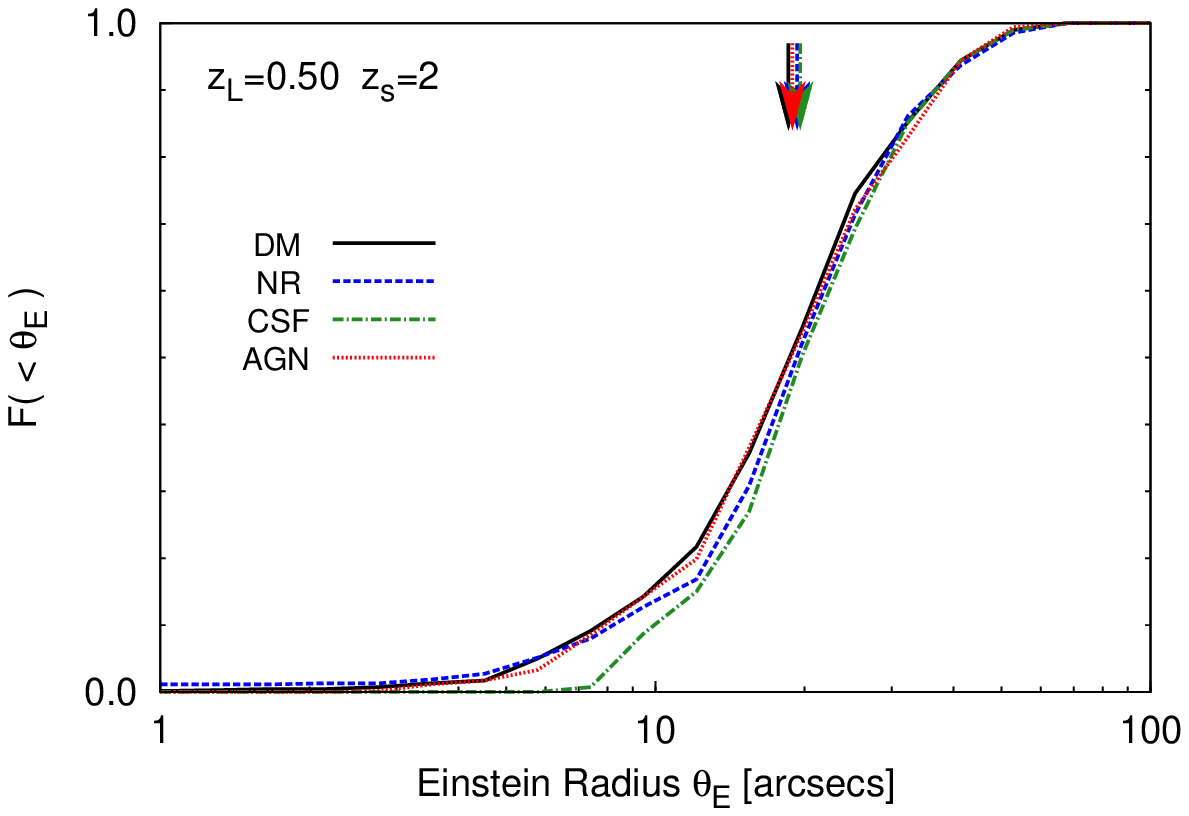}
	\end{flushright}
	\end{minipage}
	\caption{The cumulative distribution of Einstein radii
          obtained by combining 50 lines of sight
          through each of the relaxed clusters in the sample. 
          The source redshift is assumed here to be  $z_{s}=2$. On the left
          panel, we show the results for the 24 relaxed clusters with
          $\mvir>3\times10^{14}\,h^{-1}\msol$ at redshift $z_{L}=0.25$. On the right panel, we show the
          results for the 18 relaxed clusters above the same mass
          limit at redshift $z_{L}=0.5$}
	\label{EinsteinRadiiHistos}
\end{figure*}
We plot the cumulative probability distribution for Einstein radii in Figure
\ref{EinsteinRadiiHistos} combining all 50 lines of sight through each
of the relaxed clusters.
As in Section \ref{baryonsXsec}, we conduct a two-sample
Kolmogorov-Smirnov test to compare the probability distribution for each
simulation against the {\tt DM} simulation.  
We want to determine the probability, $p_{\theta}$, that the D-statistic would be atleast as large as that measured assuming the samples are drawn from the same underlying distribution. As noted before, while the clusters are independent, the many lines of sight analysed for each cluster are not
independent from each other. However, for the simulated sample, scatter due to orientation has to be
taken into account. Stacking the results for all lines of sight
analysed, we calculate the D-statistic. Defining the `sample size'
as the total number of lines of sight (50 times the number of
clusters) provides a lower limit for the p-value,
$p_{\theta}^{min}$. Alternatively, by letting the number of clusters
be the `sample size', we can obtain an upper limit on the p-value,
$p_{\theta}^{max}$. Both are listed in Table 1.

Table 1 also includes the typical Einstein radii for clusters in each
simulation, for low and high source redshifts: $z_{s}=1$ and
$z_{s}=2$. These are calculated by taking the median value of the
Einstein radius over the 50 lines of sight analysed for each clusters,
then averaging over all clusters. 
The strong lensing efficiency of a cluster is sensitive to the enclosed mass within $\EinRad$. Therefore, lens-source configurations that produce larger Einstein radii are less susceptible to the effects of baryons, since AGN lead to the re-distribution of mass within reasonable small radii; we leave a more detailed discussion to  Section \ref{profiles}, . 
Comparing the results for {\tt DM} and {\tt NR} simulations, we find that Einstein radii are similar whether the cluster lens is simulated with collision-less particles only, or with additional non-radiative gas.
Comparing the results for {\tt DM} and {\tt CSF} simulation, gas
cooling and star formation increase the predicted Einstein radii by
$\sim10$ per cent for high source redshifts ($z_{s}=2$) and
$\sim20$--40 per cent for lower source redshifts ($z_{s}=1$). This effect is more significant for lower source redshifts, for which the Einstein radii are typically smaller. 
With the inclusion of AGN feedback, the typical Einstein radii are
reduced with respect to the {\tt CSF} case. For Einstein radii calculated for source redshifts of
$z_{s}=2$, the fractional increase relative to the clusters in the
{\tt DM} simulations is only $\sim5$ per cent. This is the source
redshift that reconstructed critical curves are commonly scaled to in
observational studies. The greatest difference between predicted
Einstein radii from {\tt DM} and {\tt AGN} simulations occurs for
high-z clusters and low-z sources, which corresponds to the smaller
Einstein radii ($\theta < 20$").

%
\begin{figure*}
      \begin{minipage}{0.495\textwidth}
	\begin{flushleft}
	 \includegraphics[width=0.99\linewidth]{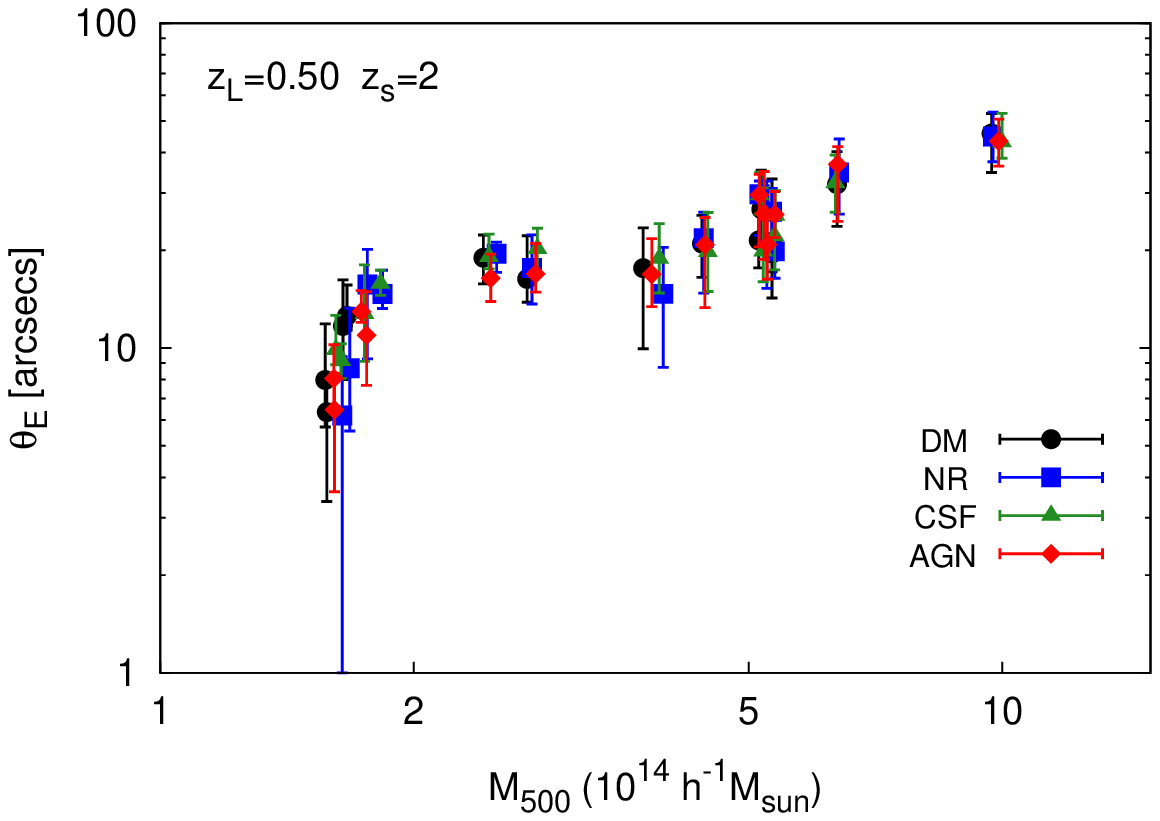}
	\end{flushleft}
      \end{minipage}
      \hfill
      \begin{minipage}{0.495\textwidth}
	\begin{flushright}
	 \includegraphics[width=0.99\linewidth]{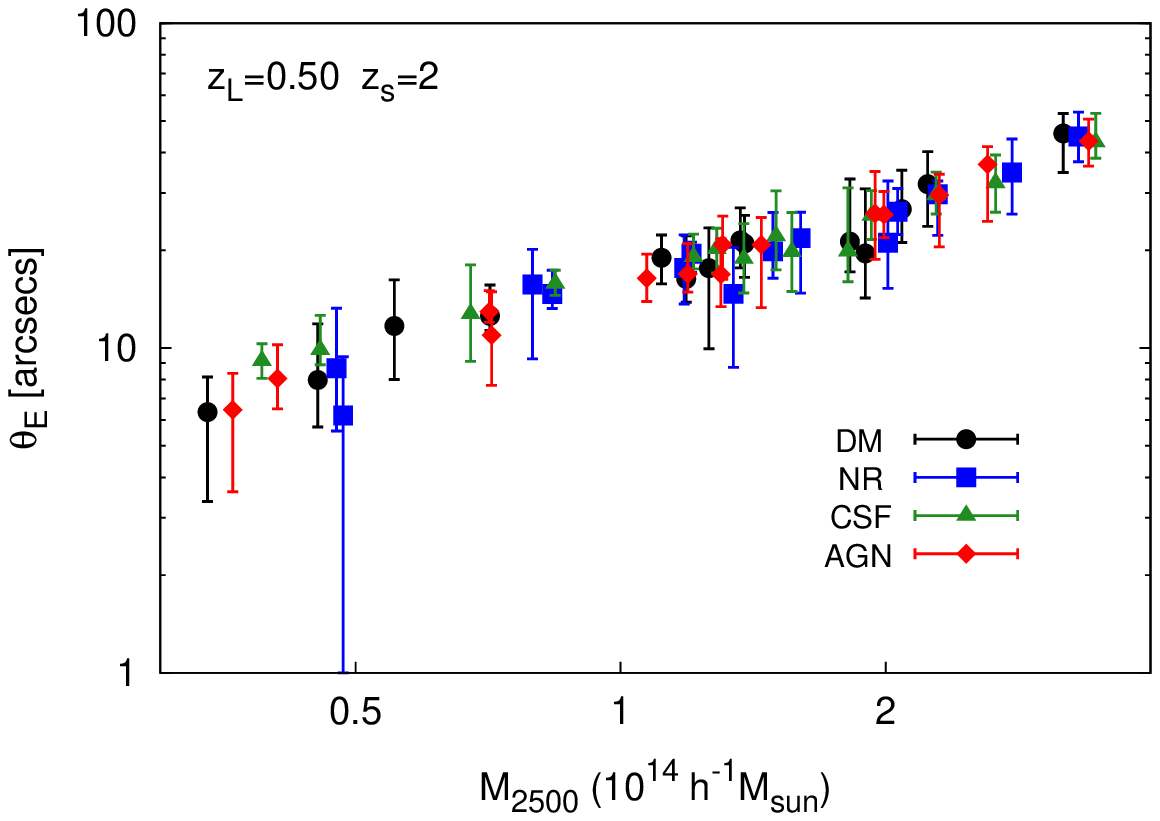}
	\end{flushright}
      	\end{minipage}
      	\caption{The Einstein radius for a source redshift of
          $z_{s}=2$ versus cluster mass combining 50 lines of sight
          through the relaxed clusters at $z_{L}=0.5$. For the left-hand panels, the
          characteristic mass is M$_{500}$ while for the right-hand panels,
          the characteristic mass is M$_{2500}$}
	\label{erVmass}
\end{figure*}
Figure \ref{erVmass} shows the Einstein radius for each cluster as a
function of its mass. We plot the median value of $\EinRad$ measured
over the 50 lines of sight for an individual cluster; the error bars,
denoting the 16th and 84th percentiles, reflect a measure of spread
associated to the line-of-sight variance. We note that one of the
clusters has a unusually large scatter in $\theta_{E}$ associated with
the different orientations. 
This cluster has a small Einstein radius and is probably on a similar
scale to angular resolution.
The cluster mass is measured at two different overdensities:
$\Delta=500$ and $\Delta=2500$.  The Einstein radius increases with
cluster mass, as expected, but the correlation is tighter at higher
overdensities. This is in line with the expectation that high
overdensities are responsible for strong lensing. The Einstein radius
for a fixed value of M$_{500}$ is slightly larger for clusters in the
{\tt CSF} simulations due to the higher concentration of these
clusters. The asphericity, reflected in the size of the error bars, is
also smaller for clusters in these simulations.

\subsection{Unrelaxed Clusters}\label{unrelaxed}
Unrelaxed clusters produce highly complex caustic structures which are
the consequence of high mass subhaloes that lie within the field of
view. These can create non-trivial complications in measurements of
both the giant arc cross-section as well as the Einstein radius. In
the previous sections we have discussed the lensing properties of
relaxed clusters so as not to be susceptible to the effects of massive
substructures. We now consider the subsample of unrelaxed clusters to
verify how their properties differ and how they would affect lensing
predictions. We remind the reader that at $z_{L}=0.25$ there are 25
unrelaxed clusters and at $z_{L}=0.5$ there are 20 unrelaxed clusters
in our sample.
%
\begin{figure}
      	\includegraphics[trim=3mm 2mm 3mm 2mm, clip, width=0.99\linewidth]{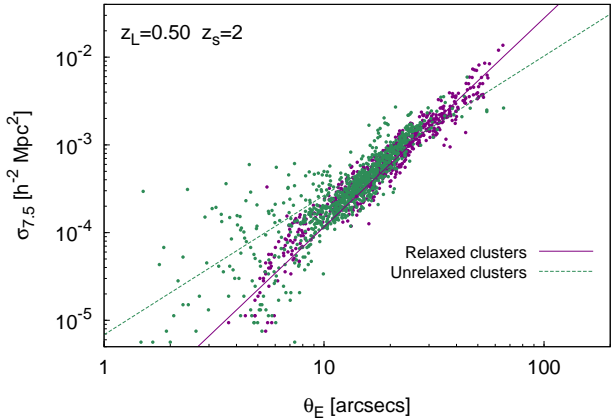}
      	\caption{The relationship between the giant-arc cross section,
          $\sigma_{7.5}$, and the Einstein radius, $\EinRad$, for each
          of the 50 lines of sight analysed for each relaxed cluster
          in the {\tt DM} simulation at $z_{L}=0.5$, for a source
          redshift of $z_{s}=2$. Results for relaxed clusters are
          shown in purple (solid line of best fit), while the
          unrelaxed clusters are shown in green (dotted line of best
          fit). The unrelaxed sub-sample exhibits a larger intrinsic
          scatter.}
	\label{CompareRelax}
\end{figure}
Figure \ref{CompareRelax} shows the correlation between $\sigma_{7.5}$
and $\EinRad$ for clusters in the {\tt DM} simulation comparing the
relaxed and unrelaxed subsample. The unrelaxed clusters in our sample
would introduce a large scatter. We verifies that, as the threshold
for the unrelaxedness parameter $\smax$, is increased, the scatter
increases in the relaxed sub-sample. This is primarily due to giant
arcs associated with substructure which induce a positive bias on the
calculation of $\sigma_{7.5}$.
The Pearson correlation coefficient, $r$, for unrelaxed clusters in
{\it all} simulations, ranges from 0.87 to 0.92 at $z_{L}=0.25$ and
ranges from 0.82 to 0.90 at $z_{L}=0.5$. 

The scatter above the $\sigma_{7.5}$-$\EinRad$ lines of best fit is
primarily because of substructures that are large enough to produce
distinct caustics (see top row of Figure \ref{SubsCaus}), which induce
a positive bias on the cross-section; as long as the caustics
associated with projected substructures are well separated from the
primary caustic, the Einstein radius measurement is not affected. On
the other hand, if substructure are projected near the clusters
centre, their caustics merge and produce highly elongated, critical
curves (see bottom row of Figure \ref{SubsCaus}), which artificially
increases both measurements of strong lensing efficiency. However, the
Einstein radius is affected more than the cross-section, 
thus down-scattering results for unrelaxed clusters with respect to
the $\sigma_{7.5}$-$\EinRad$ lines of best fit obtained for relaxed
clusters.

\begin{figure*}
	\begin{minipage}{0.495\textwidth}
	\begin{flushleft}
      	\includegraphics[width=0.99\linewidth]{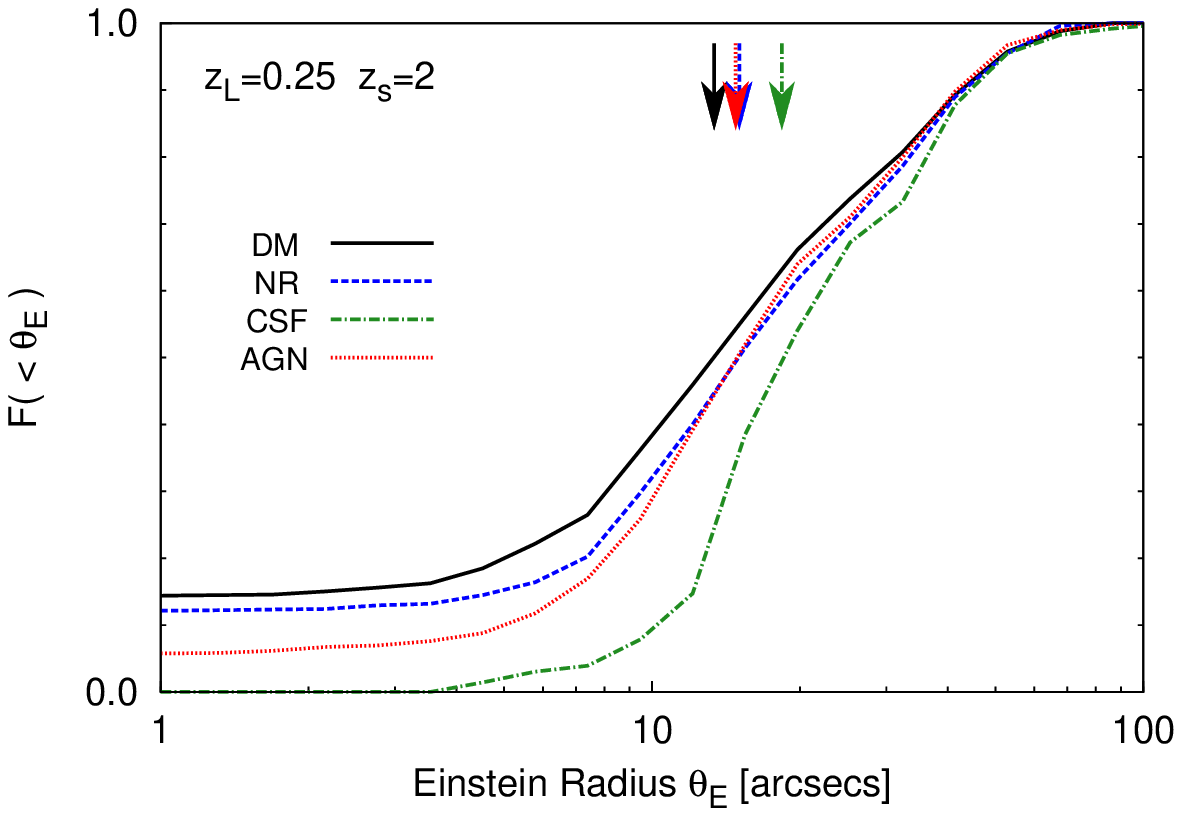}
	\end{flushleft}
	\end{minipage}
	\hfill
	\begin{minipage}{0.495\textwidth}
	\begin{flushright}
      	\includegraphics[width=0.99\linewidth]{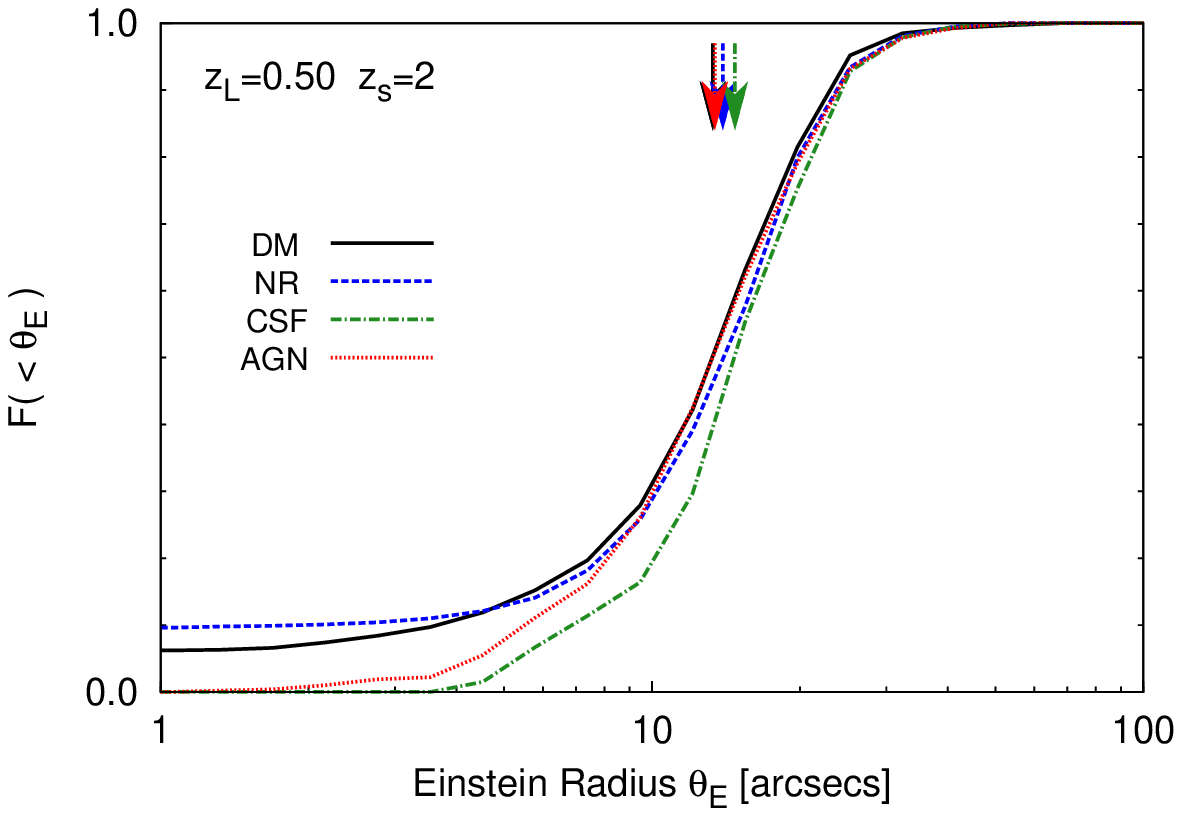}
	\end{flushright}
	\end{minipage}
	\caption{The probability distribution for Einstein radii
          combining 50 lines of sight through each of the unrelaxed
          clusters in the sample. The source redshift is $z_{s}=2$. On
          the left panel, we show the results for clusters at
          $z_{L}=0.25$; on the right panel, we show the results for
          clusters at redshift $z_{L}=0.5$}
	\label{EinsteinRadiiHistosUnrelax}
\end{figure*}
Figure \ref{EinsteinRadiiHistosUnrelax} shows the cumulative
probability distribution of Einstein radii for the unrelaxed
sub-sample. Clusters identified as unrelaxed are more likely to be
merging systems, in which the main lens has a significantly lower mass
than the total mass within the system. Hence, we find that typical
Einstein radii are only 50 -- 70 per cent of the size of Einstein radii for the relaxed counterparts. 

We may now consider whether the effect of baryons on strong lensing
properties is similar for the relaxed and unrelaxed sub-samples. Gas
cooling and star formation produce clusters with a higher mass
concentration; unrelaxed clusters have more substructures, which are
also more compact than their collisionless counterparts (see, for
example, the substructures visible in the cluster shown in Figure
\ref{CompareFourPhysics}). Therefore the unrelaxed clusters in the
{\tt CSF} simulations have larger primary caustics than their {\tt DM}
counterparts, but, compared to relaxed clusters, are {\it also} more
likely to host secondary caustics that merge with the primary and
boost the Einstein radius. Thus, the Einstein radii at $z_{s}=2$ for
clusters in the {\tt CSF} simulations are higher by $\sim20$ per cent
than for their collisionless counterparts. This boost increases to
$\sim50$--70 per cent when we measure the Einstein radii at lower
source redshifts. This is significantly larger than the boost for
relaxed clusters.

The introduction of AGN feedback once again reduces the strong lensing
efficiency of unrelaxed clusters; the Einstein radii at $z_{s}=2$ for
unrelaxed clusters in the {\tt AGN} simulations are only $5$ per cent larger than those for their collisionless counterparts in the {\tt DM} simulations. The difference in strong lensing efficiency is essentially the same for both relaxed and unrelaxed clusters. However, a mixed sample --- with both relaxed and unrelaxed clusters --- would have smaller Einstein radii on average, than a relaxed-only sample.

\subsection{Cluster Mass Profiles}\label{profiles}
In the previous sections, we have seen that gas cooling, star
formation and stellar and AGN feedback have varying and often counteracting effects on strong lensing efficiencies. Here, we look at
the mass profiles of the cluster-haloes to trace the origin of the
change in strong lensing properties. However, a detailed analysis of
the mass profile fitting and the effect of baryons on the
concentration-mass relation is beyond the scope of this paper.

\begin{figure*}
	\begin{minipage}{0.49\textwidth}
	\begin{flushleft}       	
      	\includegraphics[width=0.98\linewidth]{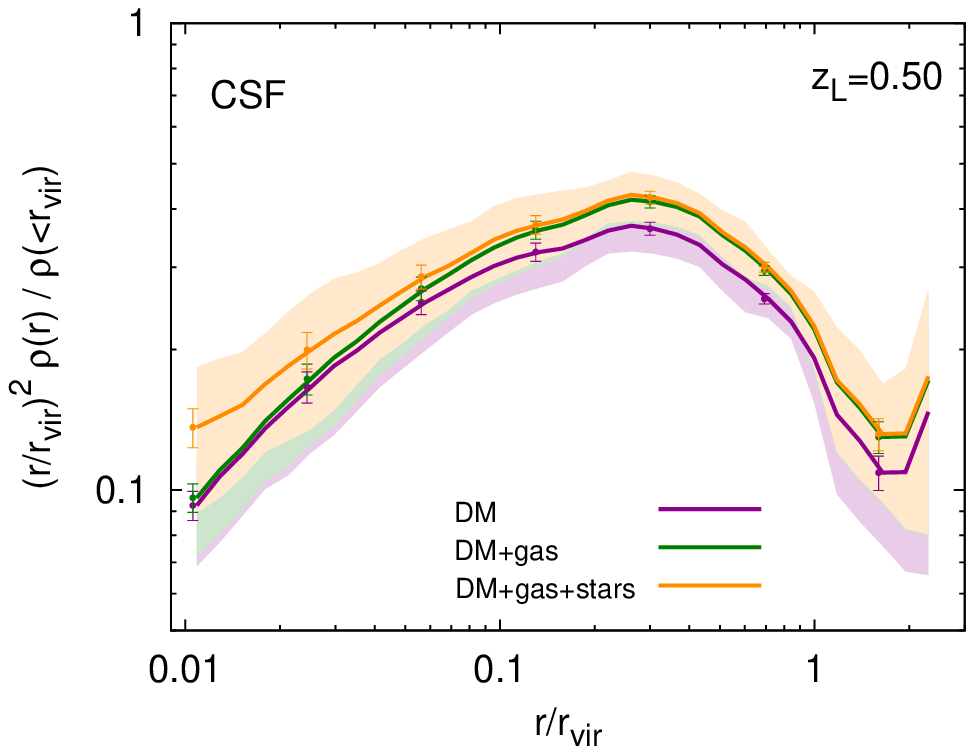}
	\end{flushleft}
	\end{minipage}
	\begin{minipage}{0.49\textwidth}
	\begin{flushright}
      	\includegraphics[width=0.98\linewidth]{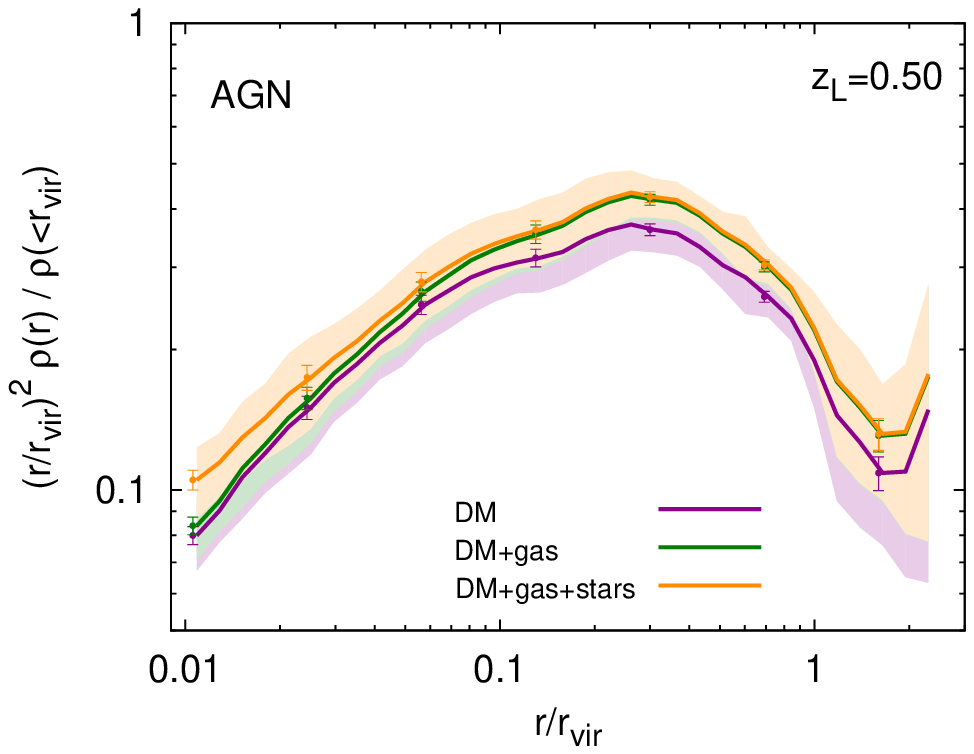}
		\end{flushright}
	\end{minipage}
      	\caption{The differential density profile of dark matter
          (purple); dark matter and gas (green); dark matter, gas and
          stars (yellow). Clusters included in analysis are the
          relaxed sub-sample at $z_{L}=0.5$ in the {\tt CSF}
          simulation (left panel) and the {\tt AGN} simulation (right
          panel). The solid lines represent the average over all
          clusters, the shaded regions (in the respective colours) are
          between the 16th and 84th percentiles among the clusters,
          while the error-bars correspond to the estimated error of
          the mean. The profile for each clusters has been normalised
          to the average mass density within the virial radius, and
          scaled with the square of the radius in order to reduce the
          dynamic range}
	\label{densprofile}
\end{figure*}
Previous analyses of non-radiative simulations in the literature have
found that the gas profiles exhibit cores
\citep[e.g.][]{R04,RZK08}. Figure \ref{densprofile} shows the
spherically averaged differential density profile for the relaxed
clusters in two of our simulations: {\tt CSF} and {\tt AGN}. The
profiles are shown for different components: dark matter, dark matter
and gas, dark matter and gas and stars. The shaded regions mark the
cluster-to-cluster scatter, while the error-bars mark the error of the
mean; 
the small size of the error bars reflect the consistency in the profiles across the clusters.

In keeping with previous analyses, we confirm that within $r \lesssim 0.1
\rvir$ gas cooling and star formation steepen the total mass profile with respect to DM--only simulations. 
In both {\tt CSF} and {\tt AGN} simulations, the baryonic component is
dominated by stars within $r \lesssim 0.1 \rvir$; the gas component
has a core and is distributed primarily beyond $10$ per cent of the
virial radius. The steepening of the inner profile is associated with
the build-up of stellar mass; AGN heating regulates star formation and
therefore reduces the inner slope. 
 We have mentioned earlier that comparisons between these two simulations must be interepreted carefully given that in the {\tt AGN} simulations SN wind-speeds are increased while simultaneously introducing AGN feedback. Interestingly, \citet{D10} have shown that in simulations without AGN feedback, the inner slope of the DM profile of clusters at $z=0$ actually steepens slightly when the SN wind-speed is reduced, but the stellar fraction within $r<0.05\rvir$ is increased significantly, which we might expect to lead to an overall increase in the inner slope of the total mass profile. They find that the introduction of AGN feedback reduces both the stellar fraction within $r<0.05\rvir$ and the inner-slope of the DM profile.
Our findings are in agreement with \citet{D10} as well as several previous studies \citep[e.g.][]{G04,P05,Mead10,Martizzi12}, in which the simulated clusters are analysed at $ z \lesssim 0.3 $.
We included in our analysis also clusters identified at higher redshift,
$z=0.5$, and found no strong redshift dependence of the results.

In order to see how the re-distribution of baryons affects the strong lensing, it is helpful to analyse the cumulative mass within different radii.
%
\begin{figure*}
	\begin{minipage}{0.49\textwidth}
	\begin{flushleft}  
      	\includegraphics[width=0.98\linewidth]{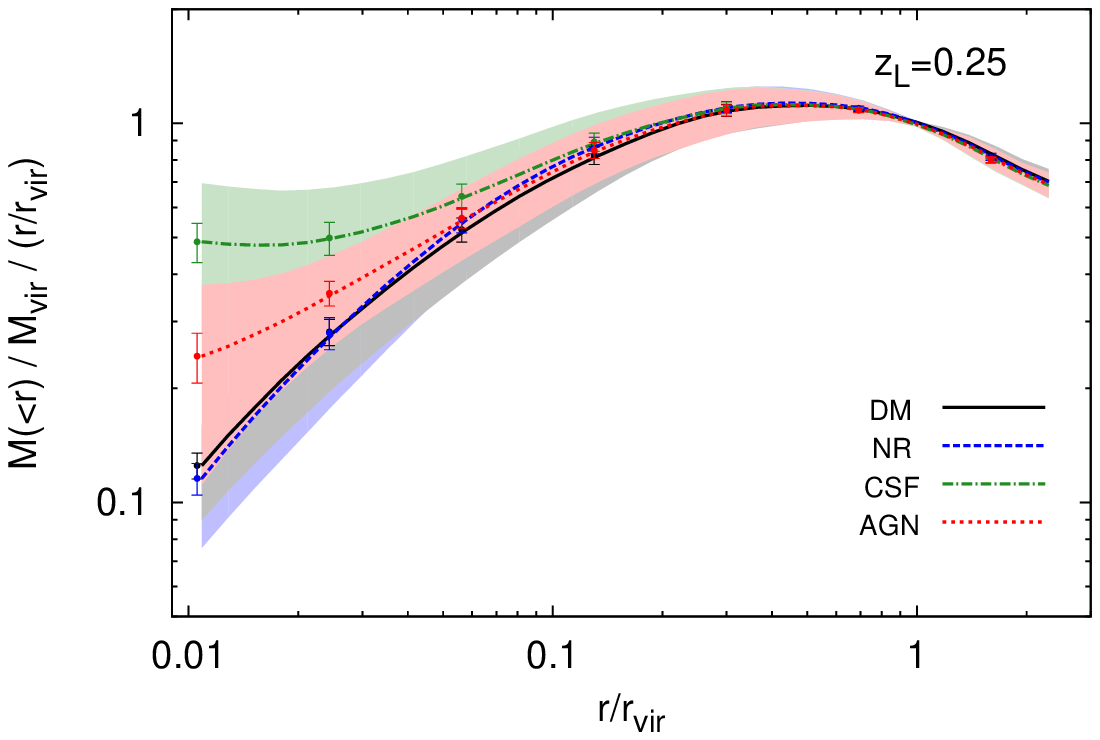}
	\end{flushleft}
	\end{minipage}
	\begin{minipage}{0.49\textwidth}
	\begin{flushright}
      	\includegraphics[width=0.98\linewidth]{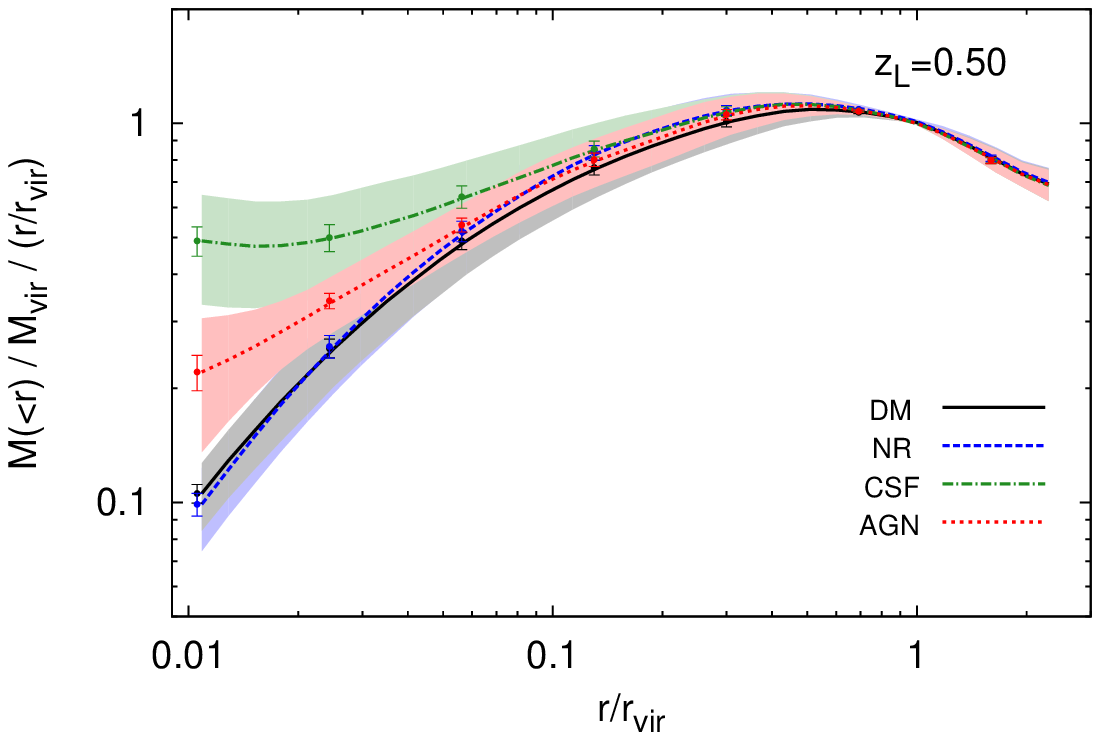} 
	\end{flushright}
	\end{minipage}
      	\caption{ The cumulative mass profile for the relaxed clusters
          at $z_{L}=0.25$ ({\it left panel}) and $z_{L}=0.5$ ({\it
            right panel}). The lines reflect the average profile
          stacking all the clusters while the shaded regions are
          between the 16th and 84th percentiles among the
          clusters. The profile for each cluster has been normalised
          to the virial mass, and scaled with the radial distance in
          order to reduce the dynamic range}
	\label{cumulmassprofile}
\end{figure*}
In Figure \ref{cumulmassprofile} we show the cumulative mass profiles
for the clusters at $z_{L}=0.25$ and $z_{L}=0.5$, respectively. The
smallest radius shown corresponds to 1 per cent of the virial radius,
which corresponds to 16 $h^{-1}$ kpc for the smallest virial radius of
all the clusters, or at least 3 times the softening length.

Our non-radiative {\tt NR} simulations generate clusters with slightly
more mass within $r < 0.1 \rvir$ than their {\tt DM} counterparts;
however, this is too mild to produce significantly stronger lenses, in
agreement with the findings of \citet{R08} and \citet{P05}.  The {\tt
  CSF} simulations produce clusters that contain significantly more
mass within $r < 0.1 \rvir$ \citep[see also][for comparable results on
steepening of density profiles in radiative
simulations]{G04,P05}. However, these simulations suffer the
`overcooling' problem --- on average, 22\% of the baryonic mass  within $r_{500}$ in {\tt CSF} clusters is in the form of stars --- thus implying that the corresponding boost in strong lensing efficiency ought to be unrealistic. A more detailed comparison between simulation and observation results on the stellar fraction will be presented by Planelles et al. (2012, in prep). 

AGN feedback reduces the total mass within 0.1 $r_{\mathrm{vir}}$ for
clusters at both redshifts
\footnote{It is informative to consider AGN
  feedback prescriptions implemented in Adaptive Mesh Refinement (AMR)
  codes, such as those presented by \citet{T11}. They report a stellar
  mass fraction of $f_{\star}\approx0.01$ which {\it increases with
    resolution} to $f_{\star}\approx0.02$. Although the concept of
  mass resolution is not directly comparable between SPH and AMR
  codes, their results do caution that AGN feedback may be sensitive
  to resolution.}.
%
\begin{figure*}
      \begin{minipage}{0.49\textwidth}
	\begin{flushleft}
	 \includegraphics[width=0.99\linewidth]{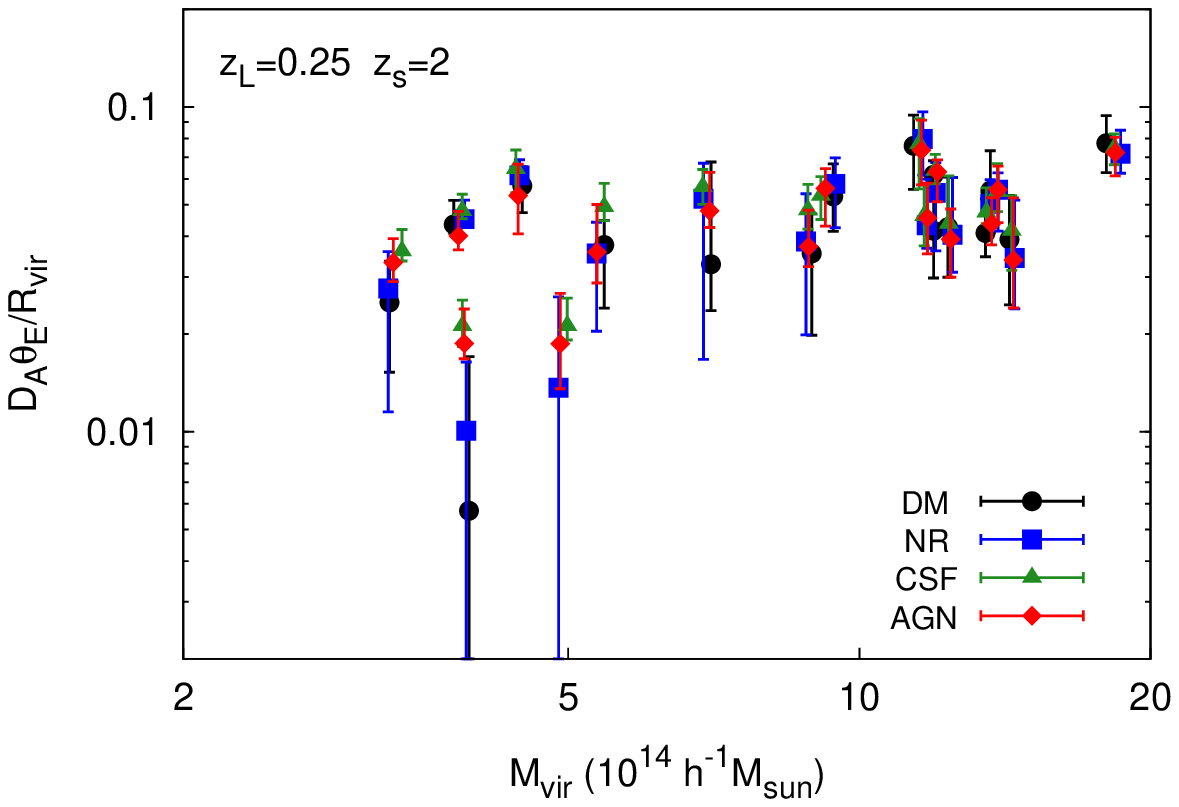}
	\end{flushleft}
      \end{minipage}
      \hfill
      \begin{minipage}{0.49\textwidth}
	\begin{flushright}
	 \includegraphics[width=0.99\linewidth]{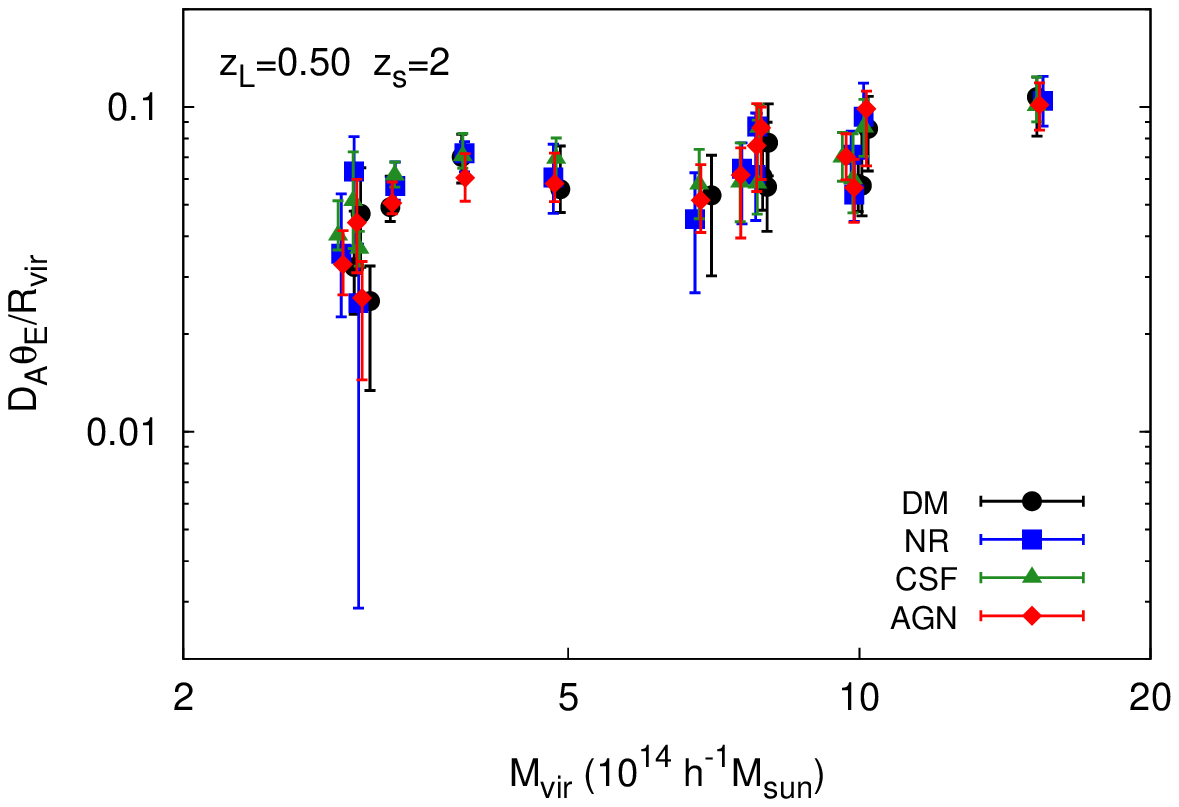}
	\end{flushright}
      	\end{minipage}
      	\caption{The Einstein radius relative to the cluster radius as
          a function of cluster mass. For each cluster, the median of
          the 50 analysed lines of sight are plotted with error bars
          reflecting the 16th and 84th percentiles over the
          distribution of lines of sight. On the left panel, we show
          relaxed clusters at $z_{L}=0.25$, while relaxed clusters at
          $z_{L}=0.5$ are on the right panel.}
	\label{erfracVmass}
\end{figure*}
If we compare the {\tt DM} and {\tt AGN} simulations, we see that
including AGN feedback results in clusters that contain more mass
within $r < 0.03 \rvir$ relative to their collision-less
counterparts. However this does not translate into larger strong lensing
efficiencies. The reason for this is seen in Figure \ref{erfracVmass} which shows the fraction of the virial radius encompassed by the Einstein radius at the cluster redshift, as a function of cluster mass. 
Typical Einstein radii are larger than the radius within which
enclosed mass is affected by baryonic physics. So while clusters in
the {\tt AGN} simulations have a different structure with respect to
their {\tt DM} counterparts, this difference does not produce stronger
lenses, for sources at $z_{s}=2$. If we consider the results for
sources at $z_{s}=1$, as shown in Table 1, the typical Einstein radii
are smaller than for sources at $z_{s}=2$. As a consequence, there is
a higher significance to the probability that clusters in the {\tt
  CSF} simulations are stronger lenses than their {\tt DM}
counterparts. Still, the strong lensing properties of clusters in the
{\tt AGN} simulations are still fairly consistent with the {\tt DM}
counterparts.

We defer a more detailed comparison with observations, extending the
simulated cluster sample to a wider range of masses, to a future
study. For the purpose of the present study, it is sufficient to say
that the suppression of star formation due to gas heating by AGN
feedback in the {\tt AGN} simulations is responsible for the decrease
in total mass in the cluster core, and subsequently, the decrease in
the strong lensing efficiency of these clusters compared to their
counterparts in the {\tt CSF} simulations.

\section{Conclusions}\label{conclude} 

In the present study, we have analysed the strong lensing efficiency
of about 15 relaxed simulated clusters at $z_{L}=0.25$ and
$z_{L}=0.5$. The clusters were re-simulated under a number of
different physical models: dark matter only ({\tt DM}); non-radiative
gas ({\tt NR}); cooling, star-formation, SN feedback ({\tt CSF}); and
additional AGN feedback ({\tt AGN}). Our main findings are as follows.
\begin{itemize}
\item There are no significant differences in the strong lensing
  properties and density profiles of clusters modelled with dark
  matter only and those that include a non-radiative gas component.
\item Gas cooling and star formation have the effect of increasing the
  giant arc counts and Einstein radii even in the presence of stellar
  feedback, which can push gas out of galaxies and reduce star
  formation.
\item When AGN feedback is included in the simulations, the strong
  lensing efficiency of the clusters are significantly
  reduced. Einstein radii at $z_{s}=2$ are only 5 per cent larger than
  those for the {\tt DM} counterparts, a boost that is increased to
  10--20 per cent for lower source redshifts. There is a smaller
  orientation dependence, reflecting a more spherical shape due to gas
  cooling, even in the presence of AGN feedback. Small Einstein radii
  ($\theta \lesssim 3$") are less likely to be produced.
\item The median Einstein radius is strongly correlated with the giant
  arc cross section for the relaxed cluster sub-sample. Unrelaxed
  clusters are especially problematic for the calculation of strong
  lensing efficiency using the production of giant lensing arcs as a
  proxy; they also have smaller strong lensing efficiencies.
\item There exists a scaling relation between strong lensing
  efficiencies and mass, particularly for higher overdensities
  ($\Delta=2500$). As we move to lower overdensities, the
  normalisation for the each simulation differs primarily due to the
  redistribution of baryons near the cluster core.
\item We search for an explanation for these findings by analysing the
  mass profiles of our cluster sample; the distribution of mass in the
  centres of clusters are significantly influenced by baryonic
  physics. AGN feedback reduces the stellar mass in the core of
  otherwise `overcooled' clusters. Nevertheless, within $r/\rvir
  \lesssim 0.03$ the total mass is still significantly larger than
  that of the {\tt DM} counterparts. However, this does not translate
  to higher lensing efficiencies for $z_{s}=2$, because $\EinRad
  \gtrsim 0.03 \rvir$ for most clusters in our chosen mass range.
\end{itemize}

Where there is overlap in cluster masses and redshifts, our results
are similar to those obtained by \citet{D10} and \citet{Mead10} with
regards to the effect of AGN feedback on density profiles and
giant-arc cross-sections. The AGN feedback prescription used in each
work are all inspired by the model of \citet{SdMH05} but with somewhat
different implementations and choice of the relevant model
parameters. Therefore, it is reassuring that the strong lensing
predictions are not sensitive to these details. However \citet{T11},
who include mechanical AGN feedback associated with jets into an Eulerian
AMR code, have noted that stellar fractions are sensitive to
implementation. An open question is then whether different
implementations of AGN feedback in different codes produce comparable
effects on the inner density profiles of massive halos and, therefore,
on the strong lensing properties of galaxy clusters.

As for the comparison with observational data, available results on
strong lensing arc statistics for samples of nearby and distant
clusters have been recently carried out, for instance, by
\cite{H10}. Furthermore, the ongoing Cluster Lensing And Supernova
survey with Hubble \citep[CLASH;][]{P12} project, thanks to the
superior sensitivity of deep HST imaging, is providing unprecedented
detail in the internal mass distribution of galaxy clusters through
strong lensing studies. These results represent an excellent test
ground for the comparison with simulation results, like those
presented in this paper. Self-consistent comparisons with data are
deferred for future study, as this will require more careful selection
of clusters, with a comparable mix of relaxed and unrelaxed objects in
the observational and simulated samples, as well as consistent methods
of measuring strong lensing properties. As \citet{WTS01} have shown, a reduction in the lensing efficiency due to a decreased mass concentration may be partially compensated by a higher magnification for individual images. Therefore, a comparison of giant arc counts will require not only a prediction for the strong lensing cross-section, but also the magnification of images that are {\it potential} giant arcs, in order to determine if they would be detected in flux-limited imaging; this is particularly important for images that lie near the flux limit.

Approaches to resolving the arc statistics problem have generally
involved questioning three aspects of the cosmological test: the
assumed source redshift distribution, the realism of simulations, and
the self-consistency in selection criteria for simulated and observed
clusters. We prefer the Einstein radius as a proxy for strong lensing,
since it allows one to avoid uncertainties with describing realistic
source populations. None of the simulated models are tuned to
reproduce {\it all} observable properties of clusters, but the {\tt
  AGN} set of simulations is the most realistic one. Our study shows
that AGN feedback is a game-changer; it's hard to explain a possible
discrepancy between observed and simulated lensing efficiency by
resorting to the effect of baryons. Therefore, it will be crucial to
make self-consistent comparisons between simulations and observations
at a range of redshifts to understand if lensing efficiency truly
presents a challenge for \lcdm.

\section*{Acknowledgments}
The authors would like to thank Volker Springel for making available
to us the non--public version of the {\small GADGET--3} code, and
Annalisa Bonafede for her help with generating the initial conditions
for the simulations. The authors would also like to thank the anonymous referee for useful comments that improved the paper, as well as Weiguang Cui, Geraint Lewis, Giuseppe Murante, Susana Planelles, Ewald Puchwein, Joop Schaye, Piero Rosati and Simon White for
helpful discussions.  Simulations have been carried out at the CINECA
supercomputing Centre in Bologna (Italy), with CPU time assigned
through ISCRA proposals and through an agreement with University of
Trieste.  MK acknowledges a fellowship from the European Commission's
Framework Programme 7, through the Marie Curie Initial Training
Network CosmoComp (PITN-GA-2009-238356). DF acknowledges funding from the Centre of Excellence for Space Sciences and Technologies SPACE-SI, an operation partly financed by the 
European Union, European Regional Development Fund and Republic of 
Slovenia, Ministry of Higher Education, Science and Technology. This work has been supported by the PRIN-INAF09 project ``Towards an Italian Network for Computational Cosmology'', by the
PRIN-MIUR09 ``Tracing the growth of structures in the Universe'' and
by the PD51 INFN grant.

\bibliography{MKpaper03}

\begin{thebibliography}{63}
\expandafter\ifx\csname natexlab\endcsname\relax\def\natexlab#1{#1}\fi

\bibitem[{{Barkana} \& {Loeb}(2010)}]{BL10}
{Barkana}, R. {Loeb}, A. 2010, \mnras, 405, 1969

\bibitem[{{Bartelmann} {et~al.}(1998){Bartelmann}, {Huss}, {Colberg},
  {Jenkins}, \& {Pearce}}]{B98}
{Bartelmann}, M., {Huss}, A., {Colberg}, J.~M., {Jenkins}, A., {Pearce}, F.~R.
  1998, \aap, 330, 1

\bibitem[{{Blumenthal} {et~al.}(1986){Blumenthal}, {Faber}, {Flores}, \&
  {Primack}}]{BFFP86}
{Blumenthal}, G.~R., {Faber}, S.~M., {Flores}, R., {Primack}, J.~R. 1986, \apj,
  301, 27

\bibitem[{{Bonafede} {et~al.}(2011){Bonafede}, {Dolag}, {Stasyszyn}, {Murante},
  \& {Borgani}}]{B11}
{Bonafede}, A., {Dolag}, K., {Stasyszyn}, F., {Murante}, G., {Borgani}, S.
  2011, \mnras, 418, 2234

\bibitem[{{Broadhurst} \& {Barkana}(2008)}]{BB08}
{Broadhurst}, T.~J. {Barkana}, R. 2008, \mnras, 390, 1647

\bibitem[{{Bryan} \& {Norman}(1998)}]{BN98}
{Bryan}, G.~L. {Norman}, M.~L. 1998, \apj, 495, 80

\bibitem[{{Bryan} {et~al.}(2012){Bryan}, {Kay}, {Duffy}, {Schaye}, {Dalla
  Vecchia}, \& {Booth}}]{Bryan12}
{Bryan}, S.~E., {Kay}, S.~T., {Duffy}, A.~R., {Schaye}, J., {Dalla Vecchia},
  C., {Booth}, C.~M. 2012, ArXiv e-prints

\bibitem[{{Chabrier}(2003)}]{chabrier03}
{Chabrier}, G. 2003, \pasp, 115, 763

\bibitem[{{Crone} {et~al.}(1996){Crone}, {Evrard}, \& {Richstone}}]{CER96}
{Crone}, M.~M., {Evrard}, A.~E., {Richstone}, D.~O. 1996, \apj, 467, 489

\bibitem[{{Cui} {et~al.}(2012){Cui}, {Borgani}, {Dolag}, {Murante}, \&
  {Tornatore}}]{Cui12}
{Cui}, W., {Borgani}, S., {Dolag}, K., {Murante}, G., {Tornatore}, L. 2012,
  \mnras, 2981

\bibitem[{{Dalal} {et~al.}(2004){Dalal}, {Holder}, \& {Hennawi}}]{DHH04}
{Dalal}, N., {Holder}, G., {Hennawi}, J.~F. 2004, \apj, 609, 50

\bibitem[{{D'Onghia} \& {Navarro}(2007)}]{DON07}
{D'Onghia}, E. {Navarro}, J.~F. 2007, \mnras, 380, L58

\bibitem[{{Duffy} {et~al.}(2008){Duffy}, {Schaye}, {Kay}, \& {Dalla
  Vecchia}}]{D08}
{Duffy}, A.~R., {Schaye}, J., {Kay}, S.~T., {Dalla Vecchia}, C. 2008, \mnras,
  390, L64

\bibitem[{{Duffy} {et~al.}(2010){Duffy}, {Schaye}, {Kay}, {Dalla Vecchia},
  {Battye}, \& {Booth}}]{D10}
{Duffy}, A.~R., {Schaye}, J., {Kay}, S.~T., {Dalla Vecchia}, C., {Battye},
  R.~A., {Booth}, C.~M. 2010, \mnras, 405, 2161

\bibitem[{{Efstathiou} {et~al.}(1990){Efstathiou}, {Sutherland}, \&
  {Maddox}}]{ESM90}
{Efstathiou}, G., {Sutherland}, W.~J., {Maddox}, S.~J. 1990, \nat, 348, 705

\bibitem[{{Fabjan} {et~al.}(2010){Fabjan}, {Borgani}, {Tornatore}, {Saro},
  {Murante}, \& {Dolag}}]{Fabjan10}
{Fabjan}, D., {Borgani}, S., {Tornatore}, L., {Saro}, A., {Murante}, G.,
  {Dolag}, K. 2010, \mnras, 401, 1670

\bibitem[{{Ferland} {et~al.}(1998){Ferland}, {Korista}, {Verner}, {Ferguson},
  {Kingdon}, \& {Verner}}]{ferland_etal98}
{Ferland}, G.~J., {Korista}, K.~T., {Verner}, D.~A., {Ferguson}, J.~W.,
  {Kingdon}, J.~B., {Verner}, E.~M. 1998, \pasp, 110, 761

\bibitem[{{Gnedin} {et~al.}(2004){Gnedin}, {Kravtsov}, {Klypin}, \&
  {Nagai}}]{G04}
{Gnedin}, O.~Y., {Kravtsov}, A.~V., {Klypin}, A.~A., {Nagai}, D. 2004, \apj,
  616, 16

\bibitem[{{Haardt} \& {Madau}(2001)}]{haardt_madau01}
{Haardt}, F. {Madau}, P. 2001, in Clusters of Galaxies and the High Redshift
  Universe Observed in X-rays, ed. D.~M. {Neumann} J.~T.~V. {Tran}

\bibitem[{{Hattori} {et~al.}(1997){Hattori}, {Watanabe}, \&
  {Yamashita}}]{HWY07}
{Hattori}, M., {Watanabe}, K., {Yamashita}, K. 1997, \aap, 319, 764

\bibitem[{{Horesh} {et~al.}(2010){Horesh}, {Maoz}, {Ebeling}, {Seidel}, \&
  {Bartelmann}}]{H10}
{Horesh}, A., {Maoz}, D., {Ebeling}, H., {Seidel}, G., {Bartelmann}, M. 2010,
  \mnras, 406, 1318

\bibitem[{{Horesh} {et~al.}(2011){Horesh}, {Maoz}, {Hilbert}, \&
  {Bartelmann}}]{H11}
{Horesh}, A., {Maoz}, D., {Hilbert}, S., {Bartelmann}, M. 2011, \mnras, 418, 54

\bibitem[{{Kazantzidis} {et~al.}(2004){Kazantzidis}, {Kravtsov}, {Zentner},
  {Allgood}, {Nagai}, \& {Moore}}]{K04}
{Kazantzidis}, S., {Kravtsov}, A.~V., {Zentner}, A.~R., {Allgood}, B., {Nagai},
  D., {Moore}, B. 2004, \apjl, 611, L73

\bibitem[{{Komatsu} {et~al.}(2011){Komatsu}, {Smith}, {Dunkley}, {Bennett},
  {Gold}, {Hinshaw}, {Jarosik}, {Larson}, {Nolta}, {Page}, {Spergel},
  {Halpern}, {Hill}, {Kogut}, {Limon}, {Meyer}, {Odegard}, {Tucker}, {Weiland},
  {Wollack}, \& {Wright}}]{K11}
{Komatsu}, E., {et~al.} 2011, \apjs, 192, 18

\bibitem[{{Kravtsov} \& {Borgani}(2012)}]{KB12}
{Kravtsov}, A. {Borgani}, S. 2012, ArXiv e-prints

\bibitem[{{Lewis} {et~al.}(2000){Lewis}, {Babul}, {Katz}, {Quinn}, {Hernquist},
  \& {Weinberg}}]{Lewis00}
{Lewis}, G.~F., {Babul}, A., {Katz}, N., {Quinn}, T., {Hernquist}, L.,
  {Weinberg}, D.~H. 2000, \apj, 536, 623

\bibitem[{{Li} {et~al.}(2005){Li}, {Mao}, {Jing}, {Bartelmann}, {Kang}, \&
  {Meneghetti}}]{L05}
{Li}, G.-L., {Mao}, S., {Jing}, Y.~P., {Bartelmann}, M., {Kang}, X.,
  {Meneghetti}, M. 2005, \apj, 635, 795

\bibitem[{{Macci{\`o}} {et~al.}(2008){Macci{\`o}}, {Dutton}, \& {van den
  Bosch}}]{M08}
{Macci{\`o}}, A.~V., {Dutton}, A.~A., {van den Bosch}, F.~C. 2008, \mnras, 391,
  1940

\bibitem[{{Martizzi} {et~al.}(2012){Martizzi}, {Teyssier}, {Moore}, \&
  {Wentz}}]{Martizzi12}
{Martizzi}, D., {Teyssier}, R., {Moore}, B., {Wentz}, T. 2012, \mnras, 422,
  3081

\bibitem[{{McCarthy} {et~al.}(2010){McCarthy}, {Schaye}, {Ponman}, {Bower},
  {Booth}, {Dalla Vecchia}, {Crain}, {Springel}, {Theuns}, \&
  {Wiersma}}]{McC10}
{McCarthy}, I.~G., {et~al.} 2010, \mnras, 406, 822

\bibitem[{{Mead} {et~al.}(2010){Mead}, {King}, {Sijacki}, {Leonard},
  {Puchwein}, \& {McCarthy}}]{Mead10}
{Mead}, J.~M.~G., {King}, L.~J., {Sijacki}, D., {Leonard}, A., {Puchwein}, E.,
  {McCarthy}, I.~G. 2010, \mnras, 406, 434

\bibitem[{{Meneghetti} {et~al.}(2000){Meneghetti}, {Bolzonella}, {Bartelmann},
  {Moscardini}, \& {Tormen}}]{M00}
{Meneghetti}, M., {Bolzonella}, M., {Bartelmann}, M., {Moscardini}, L.,
  {Tormen}, G. 2000, \mnras, 314, 338

\bibitem[{{Meneghetti} {et~al.}(2011){Meneghetti}, {Fedeli}, {Zitrin},
  {Bartelmann}, {Broadhurst}, {Gottl{\"o}ber}, {Moscardini}, \& {Yepes}}]{M11}
{Meneghetti}, M., {Fedeli}, C., {Zitrin}, A., {Bartelmann}, M., {Broadhurst},
  T., {Gottl{\"o}ber}, S., {Moscardini}, L., {Yepes}, G. 2011, \aap, 530, A17+

\bibitem[{{Meneghetti} {et~al.}(2005){Meneghetti}, {Jain}, {Bartelmann}, \&
  {Dolag}}]{M05}
{Meneghetti}, M., {Jain}, B., {Bartelmann}, M., {Dolag}, K. 2005, \mnras, 362,
  1301

\bibitem[{{Navarro} {et~al.}(1996){Navarro}, {Frenk}, \& {White}}]{NFW96}
{Navarro}, J.~F., {Frenk}, C.~S., {White}, S.~D.~M. 1996, \apj, 462, 563

\bibitem[{{Neto} {et~al.}(2007){Neto}, {Gao}, {Bett}, {Cole}, {Navarro},
  {Frenk}, {White}, {Springel}, \& {Jenkins}}]{N07}
{Neto}, A.~F., {et~al.} 2007, \mnras, 381, 1450

\bibitem[{{Oguri}(2004)}]{O04}
{Oguri}, M. 2004, PhD thesis, The University of Tokyo

\bibitem[{{Padovani} \& {Matteucci}(1993)}]{PM93}
{Padovani}, P. {Matteucci}, F. 1993, \apj, 416, 26

\bibitem[{{Postman} {et~al.}(2012){Postman}, {Coe}, {Ben{\'{\i}}tez},
  {Bradley}, {Broadhurst}, {Donahue}, {Ford}, {Graur}, {Graves}, {Jouvel},
  {Koekemoer}, {Lemze}, {Medezinski}, {Molino}, {Moustakas}, {Ogaz}, {Riess},
  {Rodney}, {Rosati}, {Umetsu}, {Zheng}, {Zitrin}, {Bartelmann}, {Bouwens},
  {Czakon}, {Golwala}, {Host}, {Infante}, {Jha}, {Jimenez-Teja}, {Kelson},
  {Lahav}, {Lazkoz}, {Maoz}, {McCully}, {Melchior}, {Meneghetti}, {Merten},
  {Moustakas}, {Nonino}, {Patel}, {Reg{\"o}s}, {Sayers}, {Seitz}, \& {Van der
  Wel}}]{P12}
{Postman}, M., {et~al.} 2012, \apjs, 199, 25

\bibitem[{{Power} {et~al.}(2011){Power}, {Knebe}, \& {Knollmann}}]{P11x}
{Power}, C., {Knebe}, A., {Knollmann}, S.~R. 2011, \mnras, 1734

\bibitem[{{Prada} {et~al.}(2012){Prada}, {Klypin}, {Cuesta}, {Betancort-Rijo},
  \& {Primack}}]{Prada12}
{Prada}, F., {Klypin}, A.~A., {Cuesta}, A.~J., {Betancort-Rijo}, J.~E.,
  {Primack}, J. 2012, \mnras, 423, 3018

\bibitem[{{Puchwein} {et~al.}(2005){Puchwein}, {Bartelmann}, {Dolag}, \&
  {Meneghetti}}]{P05}
{Puchwein}, E., {Bartelmann}, M., {Dolag}, K., {Meneghetti}, M. 2005, \aap,
  442, 405

\bibitem[{{Puchwein} \& {Hilbert}(2009)}]{PH09}
{Puchwein}, E. {Hilbert}, S. 2009, \mnras, 398, 1298

\bibitem[{{Puchwein} {et~al.}(2008){Puchwein}, {Sijacki}, \&
  {Springel}}]{PSS08}
{Puchwein}, E., {Sijacki}, D., {Springel}, V. 2008, \apjl, 687, L53

\bibitem[{{Rasia} {et~al.}(2004){Rasia}, {Tormen}, \& {Moscardini}}]{R04}
{Rasia}, E., {Tormen}, G., {Moscardini}, L. 2004, \mnras, 351, 237

\bibitem[{{Redlich} {et~al.}(2012){Redlich}, {Bartelmann}, {Waizmann}, \&
  {Fedeli}}]{R12}
{Redlich}, M., {Bartelmann}, M., {Waizmann}, J.-C., {Fedeli}, C. 2012, ArXiv
  e-prints

\bibitem[{{Rozo} {et~al.}(2008){Rozo}, {Nagai}, {Keeton}, \& {Kravtsov}}]{R08}
{Rozo}, E., {Nagai}, D., {Keeton}, C., {Kravtsov}, A. 2008, \apj, 687, 22

\bibitem[{{Rudd} {et~al.}(2008){Rudd}, {Zentner}, \& {Kravtsov}}]{RZK08}
{Rudd}, D.~H., {Zentner}, A.~R., {Kravtsov}, A.~V. 2008, \apj, 672, 19

\bibitem[{{Sadeh} \& {Rephaeli}(2008)}]{SR08}
{Sadeh}, S. {Rephaeli}, Y. 2008, \mnras, 388, 1759

\bibitem[{{Schneider} {et~al.}(1992){Schneider}, {Ehlers}, \& {Falco}}]{SEF92}
{Schneider}, P., {Ehlers}, J., {Falco}, E.~E. 1992, {Gravitational Lenses}
  (Springer-Verlag Berlin Heidelberg New York. Also Astronomy and Astrophysics
  Library)

\bibitem[{{Sijacki} {et~al.}(2007){Sijacki}, {Springel}, {Di Matteo}, \&
  {Hernquist}}]{Sijacki07}
{Sijacki}, D., {Springel}, V., {Di Matteo}, T., {Hernquist}, L. 2007, \mnras,
  380, 877

\bibitem[{{Springel}(2005)}]{S05}
{Springel}, V. 2005, \mnras, 364, 1105

\bibitem[{{Springel} {et~al.}(2005){Springel}, {Di Matteo}, \&
  {Hernquist}}]{SdMH05}
{Springel}, V., {Di Matteo}, T., {Hernquist}, L. 2005, \mnras, 361, 776

\bibitem[{{Springel} \& {Hernquist}(2003)}]{SH03}
{Springel}, V. {Hernquist}, L. 2003, \mnras, 339, 289

\bibitem[{{Teyssier} {et~al.}(2011){Teyssier}, {Moore}, {Martizzi}, {Dubois},
  \& {Mayer}}]{T11}
{Teyssier}, R., {Moore}, B., {Martizzi}, D., {Dubois}, Y., {Mayer}, L. 2011,
  \mnras, 414, 195

\bibitem[{{Thomas} {et~al.}(1998){Thomas}, {Colberg}, {Couchman}, {Efstathiou},
  {Frenk}, {Jenkins}, {Nelson}, {Hutchings}, {Peacock}, {Pearce}, \&
  {White}}]{T98}
{Thomas}, P.~A., {et~al.} 1998, \mnras, 296, 1061

\bibitem[{{Tormen} {et~al.}(1997){Tormen}, {Bouchet}, \& {White}}]{TBW97}
{Tormen}, G., {Bouchet}, F.~R., {White}, S.~D.~M. 1997, \mnras, 286, 865

\bibitem[{{Tornatore} {et~al.}(2007){Tornatore}, {Borgani}, {Dolag}, \&
  {Matteucci}}]{T07}
{Tornatore}, L., {Borgani}, S., {Dolag}, K., {Matteucci}, F. 2007, \mnras, 382,
  1050

\bibitem[{{van de Voort} {et~al.}(2011){van de Voort}, {Schaye}, {Booth}, \&
  {Dalla Vecchia}}]{vdV11}
{van de Voort}, F., {Schaye}, J., {Booth}, C.~M., {Dalla Vecchia}, C. 2011,
  \mnras, 415, 2782

\bibitem[{{White} \& {Frenk}(1991)}]{WF91}
{White}, S.~D.~M. {Frenk}, C.~S. 1991, \apj, 379, 52

\bibitem[{{Wiersma} {et~al.}(2009){Wiersma}, {Schaye}, \& {Smith}}]{WSS09}
{Wiersma}, R.~P.~C., {Schaye}, J., {Smith}, B.~D. 2009, \mnras, 393, 99

\bibitem[{{Wyithe} {et~al.}(2001){Wyithe}, {Turner}, \& {Spergel}}]{WTS01}
{Wyithe}, J.~S.~B., {Turner}, E.~L., {Spergel}, D.~N. 2001, \apj, 555, 504

\bibitem[{{Zitrin} {et~al.}(2011){Zitrin}, {Broadhurst}, {Barkana}, {Rephaeli},
  \& {Ben{\'{\i}}tez}}]{Z11a}
{Zitrin}, A., {Broadhurst}, T., {Barkana}, R., {Rephaeli}, Y.,
  {Ben{\'{\i}}tez}, N. 2011, \mnras, 410, 1939

\end{thebibliography}


\bsp

\label{lastpage}
\end{document}